\documentclass[prd,twocolumn,showpacs,nofootinbib,amsmath,amssymb]{revtex4-1}
\usepackage{graphicx,dcolumn,bm}
\usepackage{color}
\begin{document}
\title{Constraints on the annihilation cross section of dark matter
particles from anisotropies in the diffuse gamma-ray
background measured with Fermi-LAT}
\author{Shin'ichiro Ando}
%\email{s.ando@uva.nl}
\affiliation{Gravitation Astroparticle Physics Amsterdam (GRAPPA) and
Institute for Theoretical Physics, University of Amsterdam, 1090 GL
Amsterdam, The Netherlands}
\author{Eiichiro Komatsu}
\affiliation{Max-Planck-Institut f\"{u}r Astrophysik,
Karl-Schwarzschild Str. 1, 85741 Garching, Germany}
\affiliation{Kavli Institute for the Physics and
Mathematics of the Universe, Todai Institutes for Advanced Study, the
University of Tokyo, Kashiwa, Japan 277-8583 (Kavli IPMU, WPI)}
\affiliation{Texas Cosmology Center and the Department of Astronomy,
The University of Texas at Austin, 1 University Station, C1400, Austin,
TX 78712, USA}
\date{January 24, 2013}
\begin{abstract}
Annihilation of dark matter particles in cosmological halos (including
 a halo of the Milky Way) contributes to the diffuse gamma-ray background
 (DGRB). As this contribution will appear anisotropic in the sky, one
 can use the angular power spectrum of anisotropies in DGRB to constrain
 properties of dark matter particles. By comparing the updated
 analytic model of the angular power spectrum of DGRB from dark
 matter annihilation with the power spectrum recently measured
 from the 22-month data of Fermi Large Area Telescope (LAT), we place
 upper limits on the annihilation cross section of dark matter
 particles as a function of dark matter masses. We find that the current
 data exclude $\langle\sigma v\rangle\gtrsim 10^{-25}~{\rm cm^3~s^{-1}}$
 for annihilation into $b\bar{b}$ at the dark matter mass of 10~GeV,
 which is a factor of three times larger than the canonical cross
 section. The limits are weaker for larger dark matter masses. The
 limits can be improved further with more Fermi-LAT data as well as by
 using the power spectrum at lower  multipoles ($\ell\lesssim 150$),
 which are currently not used due to a potential Galactic foreground
 contamination.
\end{abstract}
\maketitle
\section{Introduction}
\label{sec:intro}
Understanding the identity and nature of dark matter, which makes up more
than 80\% of the total matter density in the Universe, is a major goal
of modern physics and cosmology. 
The most promising candidate for dark matter is the weakly interacting
massive particles (WIMPs), with which one can naturally explain the
observed dark matter density using a simple thermal freeze-out
argument~\cite{Jungman:1996}.  
If dark matter particles annihilate into standard model particles, as
expected for most WIMP scenarios, one can indirectly detect and
constrain properties of dark matter particles~\cite{Jungman:1996,
Bergstrom:2000, Bertone:2005}.
In this paper, we shall focus on high-energy (1--50~GeV) gamma-ray
photons produced by the cascade of annihilation products.

Dark matter annihilation occurs in all cosmological halos including a
halo of the Milky Way, and thus contributes to the diffuse gamma-ray background
(DGRB)~\cite{Ullio:2002, Taylor:2003, Ando:2005, Oda:2005, Pieri:2008}.
Due to the large-scale structure of the Universe, the observed gamma-ray
emission appears anisotropic in the sky in a predictable manner, which
makes it easy to identify the dark matter origin of high-energy
gamma rays in the sky~\cite{Ando:2006} (also see Refs.~\cite{Ando:2007a,
Ando:2007b, Miniati:2007, Cuoco:2007, Cuoco:2008, SiegalGaskins:2008,
Lee:2009, Taoso:2009, Fornasa:2009, SiegalGaskins:2009, Ando:2009a,
Ando:2009b, Zavala:2010, Ibarra:2010, SiegalGaskins:2011, Cuoco:2011,
Fornasa:2012}).

Recently, the Fermi-LAT collaboration has measured the power spectrum of
DGRB anisotropy from the 22-month of data~\cite{FermiAnisotropy}.
They have detected significant excesses of the angular power spectrum
over the shot noise of photons for a multipole range between $\ell =155$
and 504 and for multiple energy bins.
A further study shows that most of these excesses come from unresolved
blazars~\cite{Cuoco:2012}. 
Subtracting the estimate of the blazar contribution, we have upper
bounds on the residual anisotropy of DGRB.

In this paper, we use the upper bounds on the power spectrum to
constrain the annihilation cross section of dark matter particles.
For this purpose, we update our theoretical framework for computing
the angular power spectrum presented in Refs.~\cite{Ando:2006,
Ando:2007a, Ando:2009a} as follows: 
\begin{itemize}
\item [1.] We use the results from recent numerical simulations (e.g.,
      Ref.~\cite{Gao:2012}) to model the mass function and spatial
      distribution of subhalos within a given host halo.
\item [2.] We include contributions from both the extragalactic dark
      matter halos and the Galactic dark matter subhalos. 
\item [3.] We also include the cross correlation between dark matter
      annihilation signals and blazars. Although this term was often
      ignored in the literature (except for Ref.~\cite{Ando:2007a}),
      one should include this term for self-consistency: the same halo 
      hosting a blazar also contains annihilating dark matter particles,
      and there is a spatial correlation between halos hosting blazars
      and those not hosting blazars but containing annihilating dark
      matter particles.
\end{itemize}

This paper is organized as follows.	
In Sec.~\ref{sec:extragalactic}, we present the predicted
angular power spectrum of DGRB from extragalactic dark matter halos,
and compare this to the data to find constraints on the
annihilation cross section. 
In Sec.~\ref{sec:cross correlation}, we discuss the contribution from
the cross-correlation term with blazars.
In Sec.~\ref{sec:galactic}, we present the predicted
angular power spectrum of DGRB from Galactic subhalos, and find combined
constraints on the annihilation cross section using
extragalactic and Galactic contributions.
We conclude in Section~\ref{sec:conclusions}.
Throughout the paper, we adopt a flat cold dark matter model with a 
cosmological constant ($\Lambda$CDM) with the following cosmological
parameters: $\Omega_m = 0.277$, $\Omega_\Lambda = 0.723$, $H_0 = 100\, h
\, \mathrm{km \, s^{-1} \, Mpc^{-1}}$ with $h = 0.7$, $n_s = 0.96$, and
$\sigma_8 = 0.81$.

\section{Extragalactic contribution}
\label{sec:extragalactic}
In this section, we discuss the mean intensity and anisotropy of DGRB
from dark matter 
annihilation in extragalactic halos.
Much of the calculations are based on our earlier work~\cite{Ando:2006,
Ando:2007a}, but with extensions of the framework and significant
updates on input models as explained below.

\subsection{Mean intensity}
The mean intensity of gamma rays from dark matter
annihilation is given by\footnote{Here we define the mean intensity,
$I$, as the number of photons received per unit area, unit time, unit
energy range, and unit solid angle, i.e., $I(E) = dN / (dA dt dE
d\Omega)$.}
\begin{equation}
 I(E) = \int d\chi~W([1+z]E, \chi) \langle \delta^2 \rangle,
  \label{eq:intensity}
\end{equation}
where $E$ is the energy of photons, $\chi$ is the comoving distance
to a source at redshift $z$ [$\chi$ and $z$ are used interchangeably
through the relation, $d\chi / dz = c / H(z)$],
$\langle \delta^2 \rangle$ is the variance of the overdensity field,
$\delta = (\rho - \langle \rho \rangle ) / \langle \rho \rangle$, and 
is often also referred to as an ``intensity multiplier.''
$W(E, z)$ is the window function that contains particle-physics
information such as a velocity-averaged annihilation cross section times
relative velocity, $\langle \sigma v\rangle$, a dark matter mass,
$m_{\rm dm}$, and a gamma-ray spectrum per annihilation, $dN_{\gamma,
{\rm ann}} / dE$:
\begin{equation}
 W(E, z) = \frac{\langle \sigma v \rangle}{8\pi}
  \left(\frac{\Omega_{\rm dm} \rho_c}{m_{\rm dm}}\right)^2 (1+z)^3
  \frac{dN_{\gamma, {\rm ann}}}{dE}e^{-\tau(E, z)},
  \label{eq:window}
\end{equation}
where $\Omega_{\rm dm} = 0.23$ is the density parameter of dark matter, and
$\rho_c$ is the critical density of the present Universe. Here, $\tau(E,
z)$ is the optical depth for a gamma ray emitted at energy $E$, for
which we adopt the model of Ref.~\cite{Gilmore:2011}.
Note that the annihilation cross section required to produce dark matter
at the right relic density by the thermal freeze-out mechanism
with S-wave annihilation is $\langle \sigma v \rangle \simeq 3 \times
10^{-26} ~ \mathrm{cm^3 ~ s^{-1}}$, which is largely independent of the
dark matter mass~\cite{Jungman:1996}.

The intensity multiplier is
\begin{eqnarray}
 \langle \delta^2 \rangle & = & \left(\frac{1}{\Omega_m \rho_c}\right)^2
  \int dM \frac{dn(M, z)}{dM} [1 + b_{\rm sh}(M)]
  \nonumber \\ &&\times {}
  \int dV \rho_{\rm host}^2(r|M),
  \label{eq:multiplier}
\end{eqnarray}
where $M$ is the virial mass, $dn/dM$ is the halo mass function, for
which we adopt an ellipsoidal collapse model~\cite{Sheth:1999,
Sheth:2001}, $\rho_{\rm host}(r | M)$ is the density profile of a host
halo of mass $M$, $r$ is the comoving radius from the halo center, and
$b_{\rm sh}(M)$ is a 
boost factor due to annihilation in subhalos.

We adopt an Navarro-Frenk-White (NFW)~\cite{NFW} profile for host halos,
\begin{equation}
 \rho_{\rm host}(r) = \frac{\rho_s}{(r/r_s) (r/r_s + 1)^2},
  \label{eq:NFW}
\end{equation}
where $\rho_s$ and $r_s$ are the scale density and the scale radius,
respectively, and this relation holds out to the virial radius,
$r_{\rm vir}$, which, in turn, is given as a function of $M$ and $z$
through the relation: $M = 4\pi r_{\rm vir}^3 \Delta_{\rm vir}(z)
\rho_c(z) / 3$, with $\Delta_{\rm vir}(z) = 18\pi^2 + 82 d - 39 d^2$ and $d
= \Omega_m (1+z)^3 / [\Omega_m (1+z)^3 + \Omega_\Lambda] -
1$~\cite{Bryan:1998}.

The scale radius is defined as $r_s = r_{\rm vir} / c_{\rm vir}$,
where $c_{\rm vir}(M, z)$ is the concentration parameter, for which we
adopt the model of Ref.~\cite{Bullock:2001} for masses below $2.5 \times
10^{14} M_\odot$ and that of Ref.~\cite{Duffy:2008} otherwise.
By taking the volume integral of the density profile, $\rho_{\rm
host}(r)$, out to $r_{\rm vir}$ and equating it to $M$, we obtain the
scale density as 
\begin{equation}
 \rho_s = \frac{M}{4\pi r_s^3}
  \left[\ln (1 + c_{\rm vir}) - \frac{c_{\rm vir}}{1 + c_{\rm vir}}
  \right]^{-1}.
\end{equation}
The volume integral of the density squared has an analytic form:
\begin{equation}
 \int dV \rho_{\rm host}^2 = \frac{4\pi r_s^3 \rho_s^2}{3}
  \left[1 - \frac{1}{(1 + c_{\rm vir})^3}\right].
  \label{eq:volume integral of density squared}
\end{equation}

The gamma-ray intensity is further boosted by annihilation in subhalos,
which is represented by the boost factor, $b_{\rm sh}(M)$, for which we
adopt a fitting formula based on results of recent numerical
simulations~\cite{Gao:2012}: $b_{\rm sh} \approx 110 (M_{200} / 10^{12}
M_\odot)^{0.39}$, where $M_{200}$ is an enclosed mass within a radius
$r_{200}$ in which the average density is 200 times the critical
density; there is a simple relation between $M_{200}$ and the virial
mass $M$~\cite{Hu:2003}.
This boost is realized if the subhalo mass function extends down to
Earth-mass scales, $M_{\rm sh, min} = 10^{-6} M_\odot$, which is a
typical cutoff scale for the neutralino dark matter~\cite{Green:2004,
Green:2005, Diemand:2005, Diemand:2006}.
We note, however, that the boost factor strongly depends on the minimum
subhalo mass chosen, $b_{\rm sh} \propto M_{\rm sh,
min}^{-0.2}$~\cite{Gao:2012}.
Given that a wide range of minimum subhalo mass is still allowed, as
small as $\sim$10$^{-12} M_\odot$~\cite{Profumo:2006}, the annihilation
rate may be boosted even further.

\begin{figure}[t]
 \begin{center}
  \includegraphics[width=8.5cm]{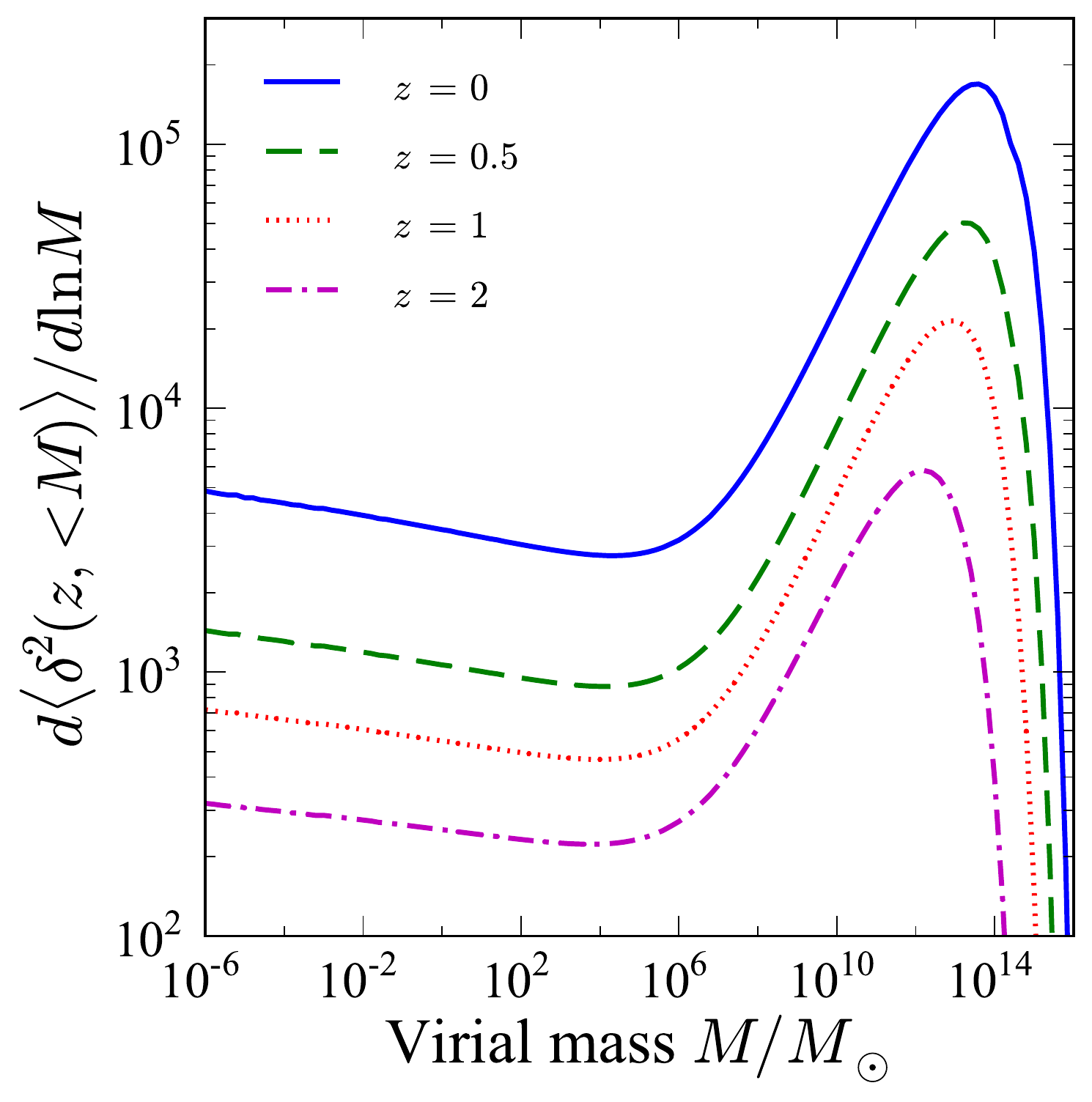}
  \caption{Contribution to the intensity multiplier, $\langle\delta^2\rangle$
  [Eq.~(\ref{eq:multiplier})], from different mass 
  ranges at various redshifts.}
  \label{fig:dfdM}
 \end{center}
\end{figure}
\begin{figure}[t]
 \begin{center}
  \includegraphics[width=8.5cm]{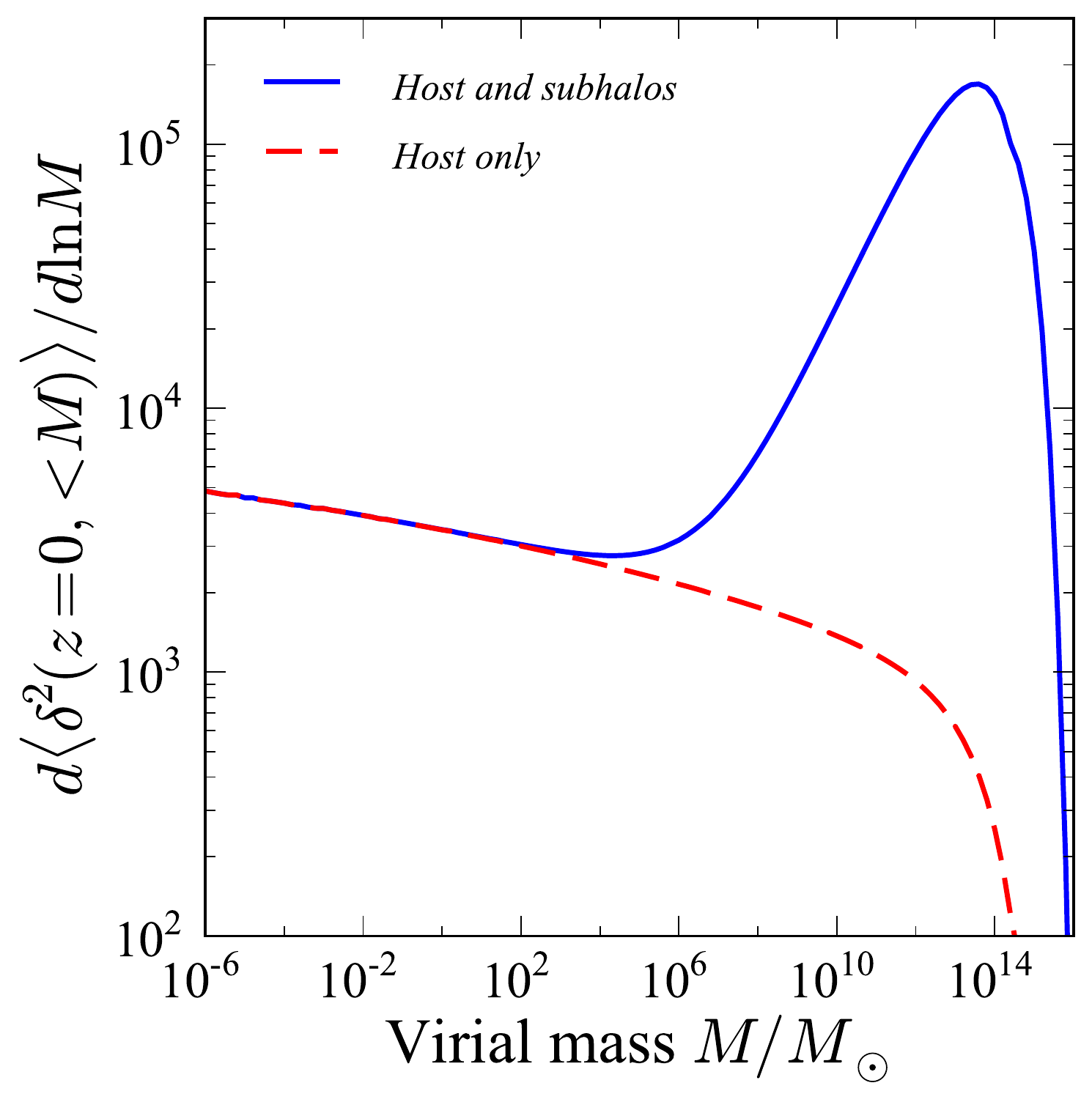}
  \caption{Contributions to the intensity multiplier,
  $\langle\delta^2\rangle$ [Eq.~(\ref{eq:multiplier})], from host halos
  (dashed) and host halos and subhalos (solid) at $z = 0$. The solid
  line is the same as that in Fig.~\ref{fig:dfdM}.}
  \label{fig:dfdM_host}
 \end{center}
\end{figure}

\begin{figure}[t]
\begin{center}
 \includegraphics[width=8.5cm]{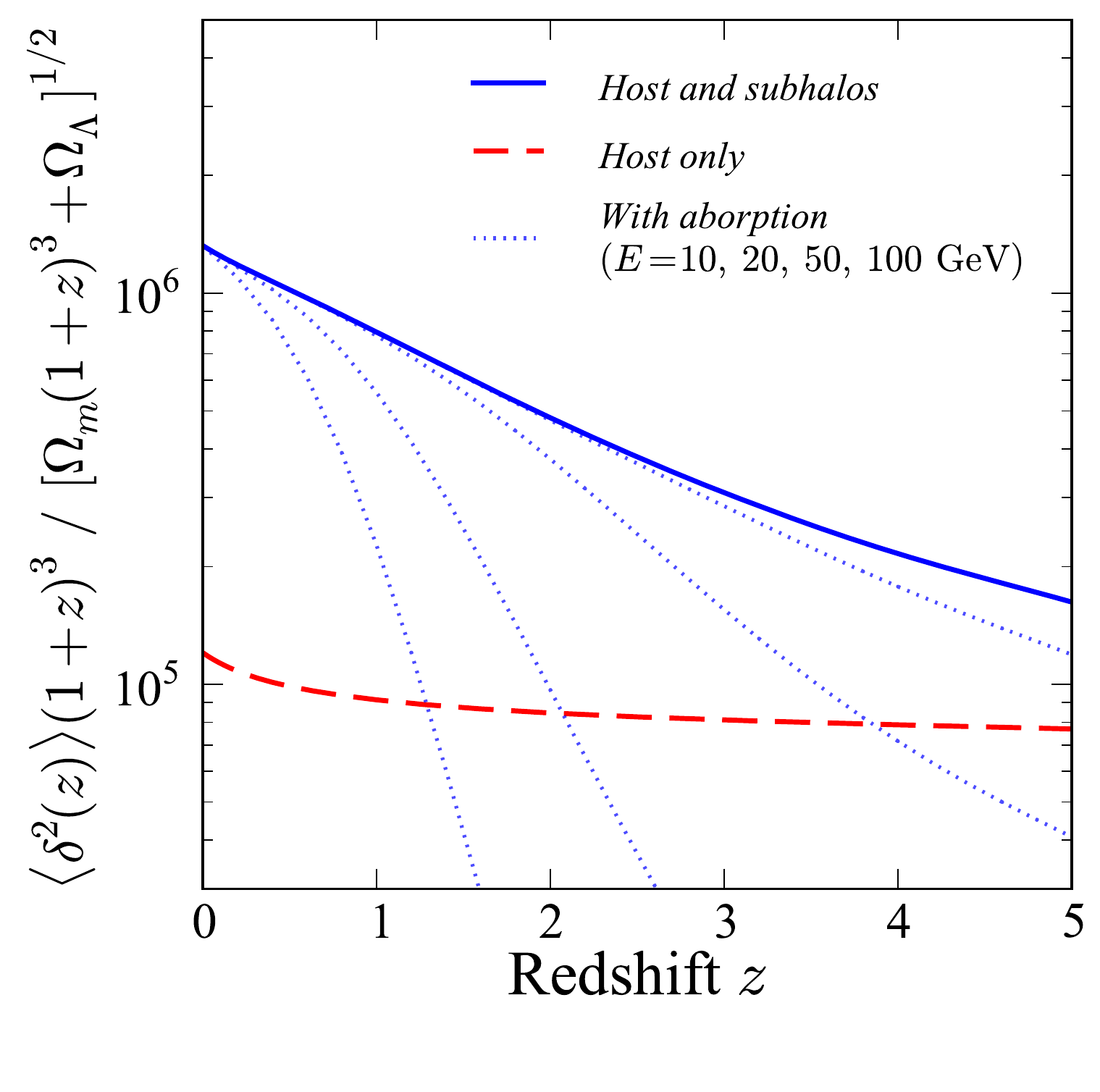}
 \caption{Intensity multiplier as a function of
 redshifts. The solid line shows the total intensity
 multiplier from host 
 halos and subhalos, while the dashed line shows the intensity
 multiplier only from host halos. The dotted lines show the total intensity
 multiplier multiplied by the absorption factors at various energies
 ($E=100$, 50, 20, and 10~GeV from left to right).}
 \label{fig:multiplier}
\end{center}
\end{figure}

\begin{figure}[t]
 \begin{center}
  \includegraphics[width=8.5cm]{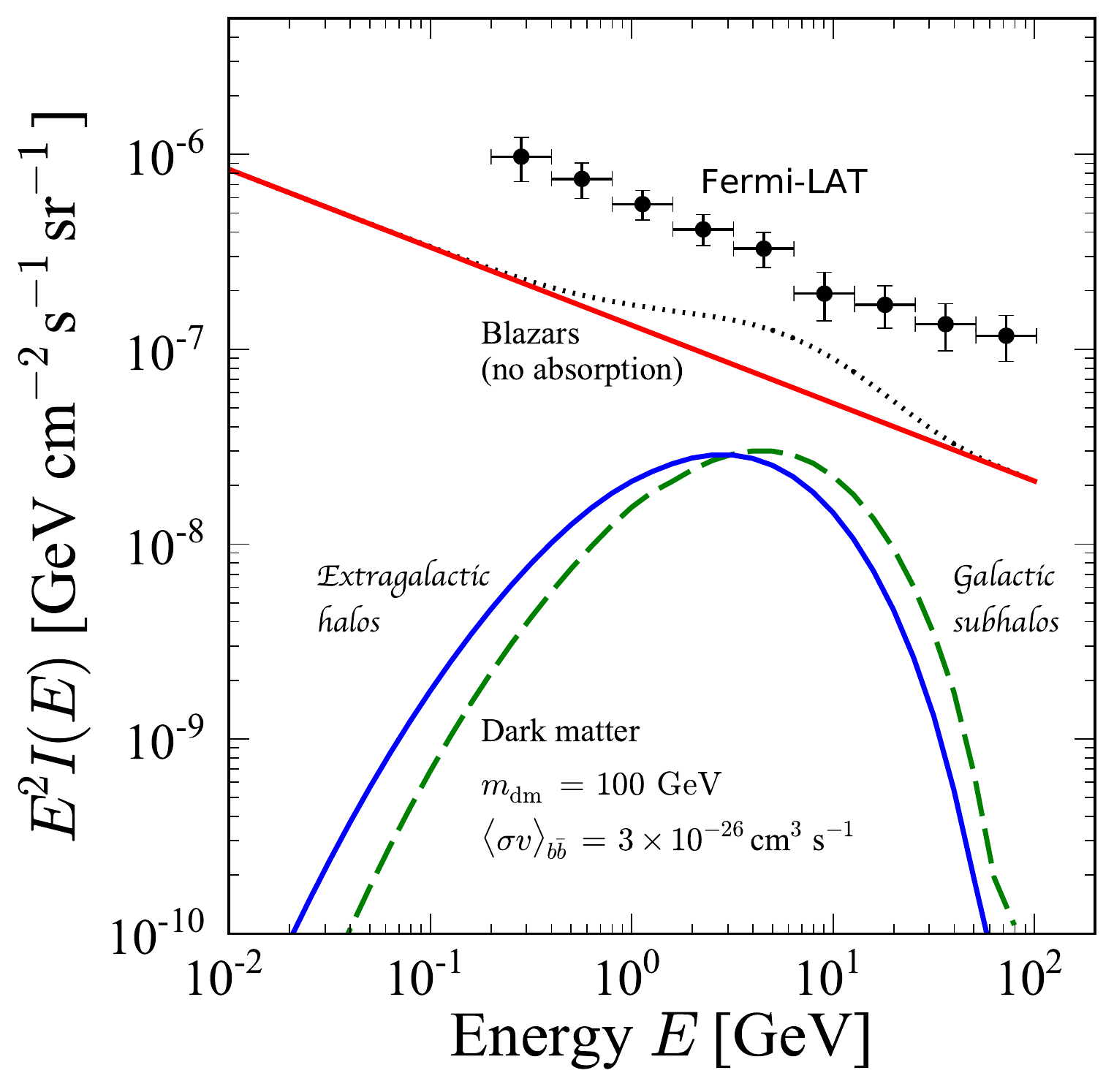}
  \caption{Predicted mean intensity of DGRB from 
  dark matter annihilation via the $b\bar b$ channel. The
  dark matter 
  mass is $m_{\rm dm} = 100\,{\rm
  GeV}$, and the cross section is at its canonical value, $\langle
  \sigma v \rangle = 3\times 10^{-26}\, 
  \mathrm{cm^3\, s^{-1}}$. The solid line shows the
  extragalactic contribution, while the dashed line shows the Galactic
  subhalo contribution (no smooth Galactic contribution is
  included).
  The Fermi-LAT data~\cite{FermiDiffuse} as
  well as the contribution from unresolved blazars are also shown for
  comparison.}
  \label{fig:spectrum}
 \end{center}
\end{figure}

Figure~\ref{fig:dfdM} shows the integrand of Eq.~(\ref{eq:multiplier})
as a function of the virial mass, $M$, for various redshifts, $z$.
Namely, this shows what fraction of $\langle \delta^2 \rangle$ is
contributed from which mass ranges.
As can be seen from this figure, the dominant contribution comes from
cluster-size halos ($M\sim 10^{14}~M_\odot$) at $z\sim 0$ and galaxy-size
halos ($M\sim 10^{12}~M_\odot$) at $z\sim 2$.
This is mainly because of the boost due to annihilation in subhalos.
In Fig.~\ref{fig:dfdM_host}, we compare the contribution
from the host halos (dashed line) and that from the host halos and
subhalos (solid line) at $z = 0$.
If there are no gamma ray from subhalos, then the mean intensity would
be dominated by the smallest dark matter halos.

Figure~\ref{fig:multiplier} shows the intensity multiplier $\langle
\delta^2 \rangle$ as a function of redshifts for the case with and
without subhalo contributions.
Here we multiply $\langle \delta^2 \rangle$ by $(1+z)^3 / \sqrt{\Omega_m
(1+z)^3 + \Omega_\Lambda}$ to show 
contributions to the mean intensity from different redshift ranges.
One can see that the presence of subhalos boosts the intensity by a
factor of 10 at low redshifts, and by a factor of $\sim 2$ even at $z =
5$.
The dotted lines are further multiplied by the
absorption factor, $e^{-\tau}$, for observed energies of $E = 10$, 20, 50,
and 100~GeV.
There is little absorption for photons received below 10~GeV, but
this effect is significant for energies above tens of GeV and should be
taken into account.

Figure~\ref{fig:spectrum} shows the predicted mean intensity of DGRB
from dark matter annihilation with $m_{\rm dm} = 100 ~ {\rm 
GeV}$, $\langle \sigma v \rangle = 3 \times 10^{-26} ~ \mathrm{cm^3 ~
s^{-1}}$, and the $b\bar b$ annihilation channel.
This model gives the dark matter contribution that is as large as 10\%
of the mean intensity measured by Fermi-LAT~\cite{FermiDiffuse} at
$E\sim 10$~GeV. 
This contribution is quite significant given that even the most
dominant contributors known to date, i.e., unresolved blazars, contribute to
DGRB at around 
the same level~\cite{FermiDiffuseSource}.

\subsection{Angular power spectrum}

The angular power spectrum at a given multipole, $\ell$, is given by
\begin{equation}
 C_\ell(E) = \int \frac{d\chi}{\chi^2} W^2([1+z]E, z)
  P_{\delta^2} \left(k=\frac{\ell}{\chi}, z\right).
  \label{eq:angular power spectrum}
\end{equation}
Following conventions of recent publications (e.g.,
Ref.~\cite{FermiAnisotropy}), our definition of the angular power
spectrum [Eq.~(\ref{eq:angular power spectrum})] has the units of
$\mathrm{(cm^{-2}~ s^{-1} ~ sr^{-1} ~ GeV^{-1})^2 ~ sr}$,\footnote{When
we compare theoretical predictions with the data, we must
integrate a gamma-ray intensity in a given direction over energy within
a given energy  bin. We do this by replacing the window function,
$W([1+z]E, z)$, in Eqs.~(\ref{eq:intensity}) and (\ref{eq:angular power
spectrum}) with the window function integrated over a given energy
range. This gives $C_\ell$ in units of $\mathrm{(cm^{-2}~ s^{-1}~
sr^{-1})^2 ~ sr}$.} which is
referred to as the {\it intensity} angular power spectrum. In order to
obtain the {\it fluctuation} angular power spectrum that has
units of sr and is adopted in the earlier papers (e.g.,
Ref.~\cite{Ando:2006}), one simply divides the intensity power spectrum
by the mean-intensity squared, $I^2(E)$. 

Here, $P_{\delta^2}(k, z)$ is the power spectrum of the overdensity
squared, $\delta^2$, which can be divided into one- and two-halo
terms~\cite{Ando:2006}:
\begin{equation}
 P_{\delta^2}(k, z) = P_{\delta^2}^{\rm 1h}(k, z) +
  P_{\delta^2}^{\rm 2h} (k, z).
\end{equation}
The one-halo term correlates two points in one identical halo, whereas
the two-halo term does that in two distinct halos.
Correspondingly, the angular power spectrum is also divided into two
terms:
\begin{equation}
 C_\ell(E) = C_\ell^{\rm 1h}(E) + C_\ell^{2h}(E).
\end{equation}
These two terms of the power spectrum can be explicitly written as
\begin{eqnarray}
 P_{\delta^2}^{\rm 1h}(k, z) &=&
  \left(\frac{1}{\Omega_m \rho_c}\right)^4 \int
  dM \frac{dn(M, z)}{dM}  |\tilde u(k|M)|^2
  \nonumber\\&&{}\times
  \left[(1 + b_{\rm sh}(M)) \int dV \rho^2_{\rm host}(r|M)\right]^2,
  \label{eq:1h} \\
 P_{\delta^2}^{\rm 2h}(k, z) &=&
  \left[
   \left(\frac{1}{\Omega_m \rho_c}\right)^2
   \int dM \frac{dn(M,z)}{dM} \tilde u(k|M)b_1(M,z)
  \right. \nonumber\\&& \times \left.
  (1 + b_{\rm sh}(M))\int dV \rho_{\rm host}^2(r|M)
  \right]^2 P_{\rm lin}(k, z),
  \nonumber\\
 \label{eq:2h}
\end{eqnarray}
where $P_{\rm lin}(k, z)$ is the linear power spectrum of the matter
density field $\delta$, and $b_1(M,
z)$ is the linear halo bias~\cite{Sheth:2001}.
The power spectrum of $\delta^2$, $P_{\delta^2}$, depends on profiles of
density {\it squared} in a halo of mass $M$, $u(r|M)$, where $u(r|M)$ is
normalized such that its volume integration becomes unity; $\tilde
u(k|M)$ is the Fourier transform of $u(r|M)$.

\begin{figure}[t]
 \begin{center}
  \includegraphics[width=8.5cm]{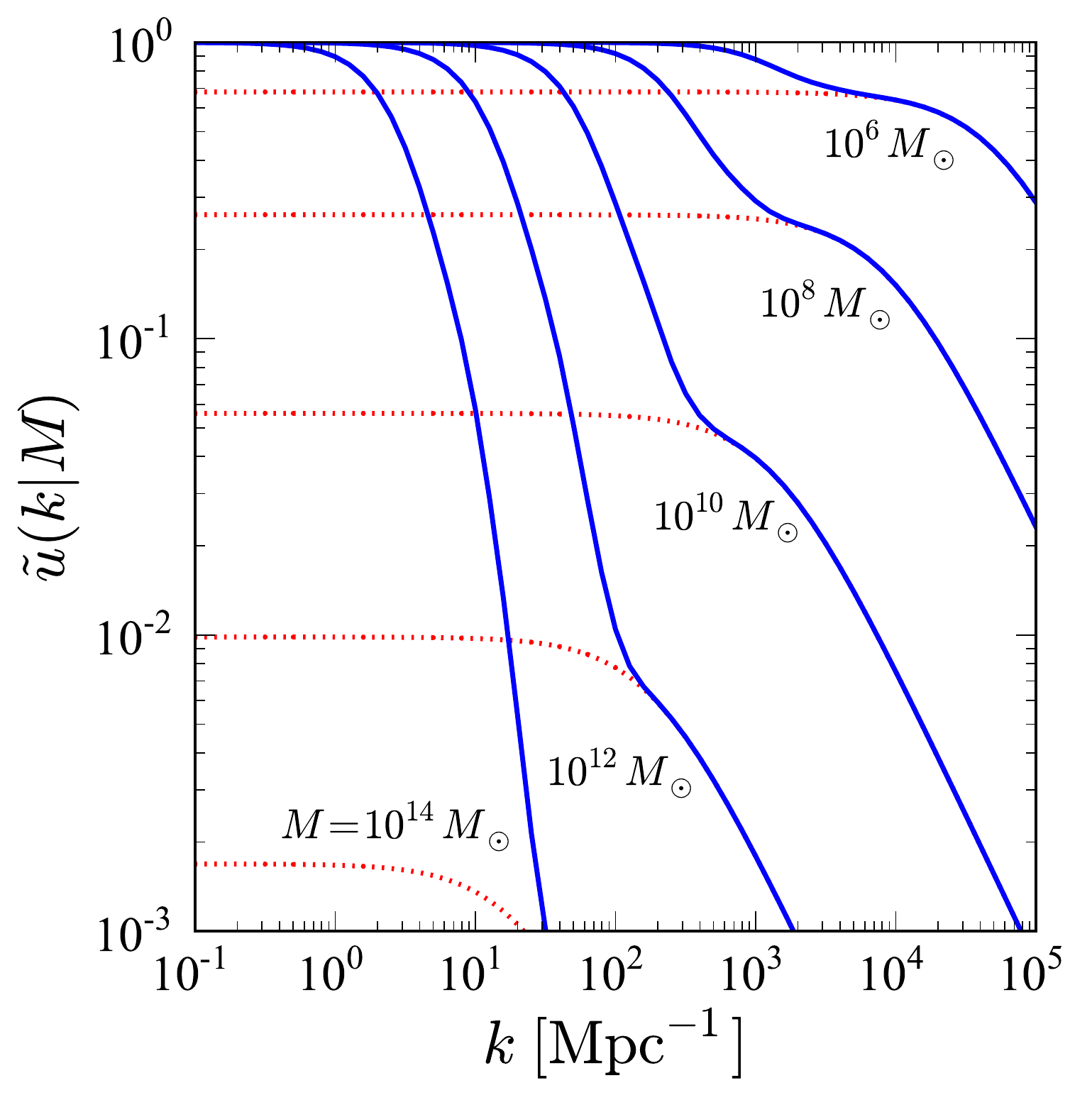}
  \caption{Fourier transform of the density-squared
  profile, $\tilde u(k|M)$, for various host-halo masses. The solid
  lines show the  total (host halo and subhalos) profile, whereas the
  dotted lines show the host-halo profiles. Note that the
  total $\tilde u(k|M)$ is normalized to unity at $k\to 0$, and thus the
  dotted lines do not approach unity at $k\to 0$ but
  approaches $1/(1+b_{\rm sh})$ [see Eq.~(\ref{eq:weightedsum})].}
  \label{fig:uFourier}
 \end{center}
\end{figure}
\begin{figure}[t]
 \begin{center}
  \includegraphics[width=8.5cm]{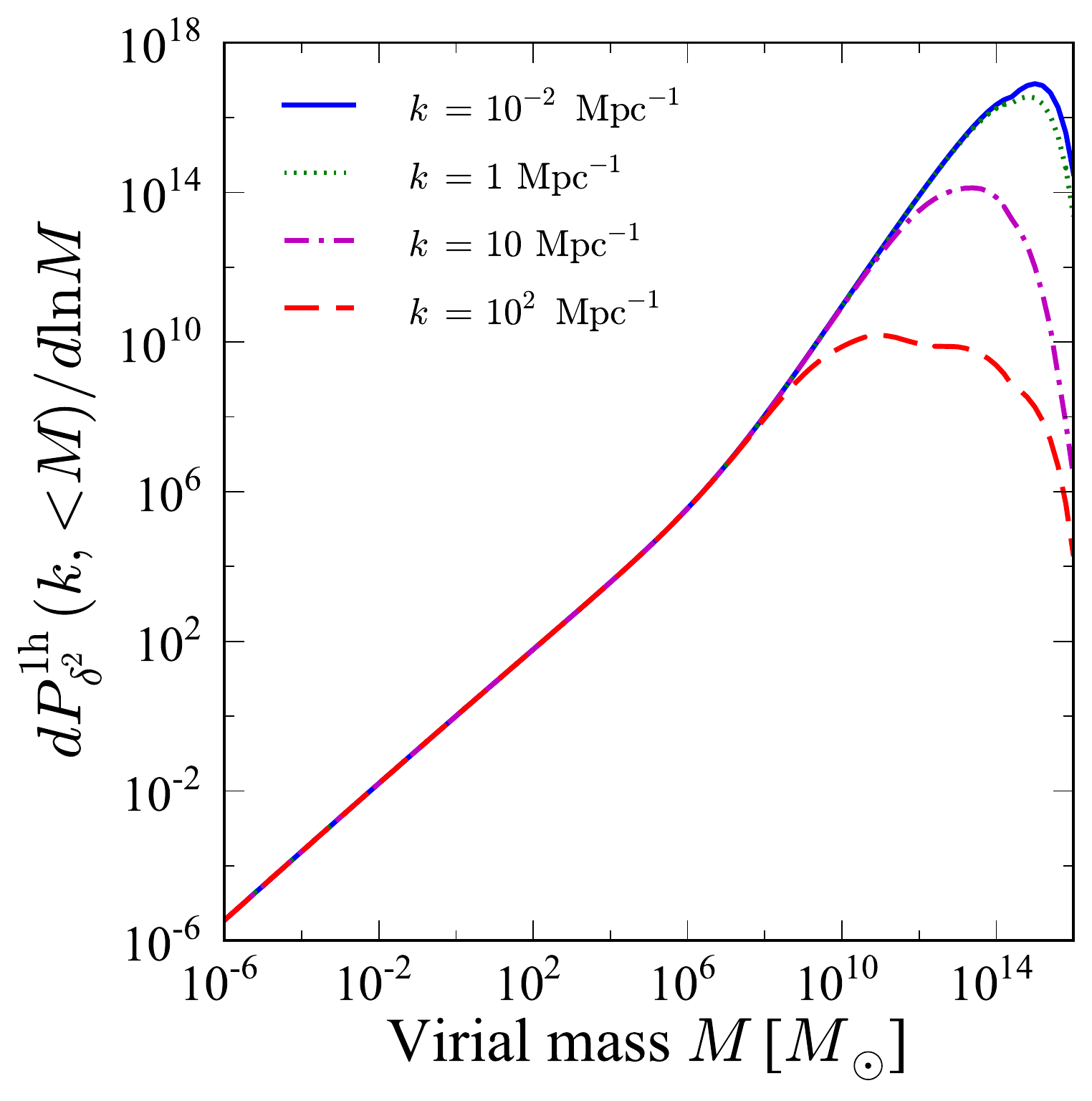}
  \caption{Relative contributions to the one-halo power spectrum
  $P_{\delta^2}^{1h}$ at $z = 0$ as a function of masses and
  wave numbers. The lines are normalized at $M = 1~M_\odot$.}
  \label{fig:dPdM}
 \end{center}
\end{figure}
\begin{figure}[t]
 \begin{center}
  \includegraphics[width=8.5cm]{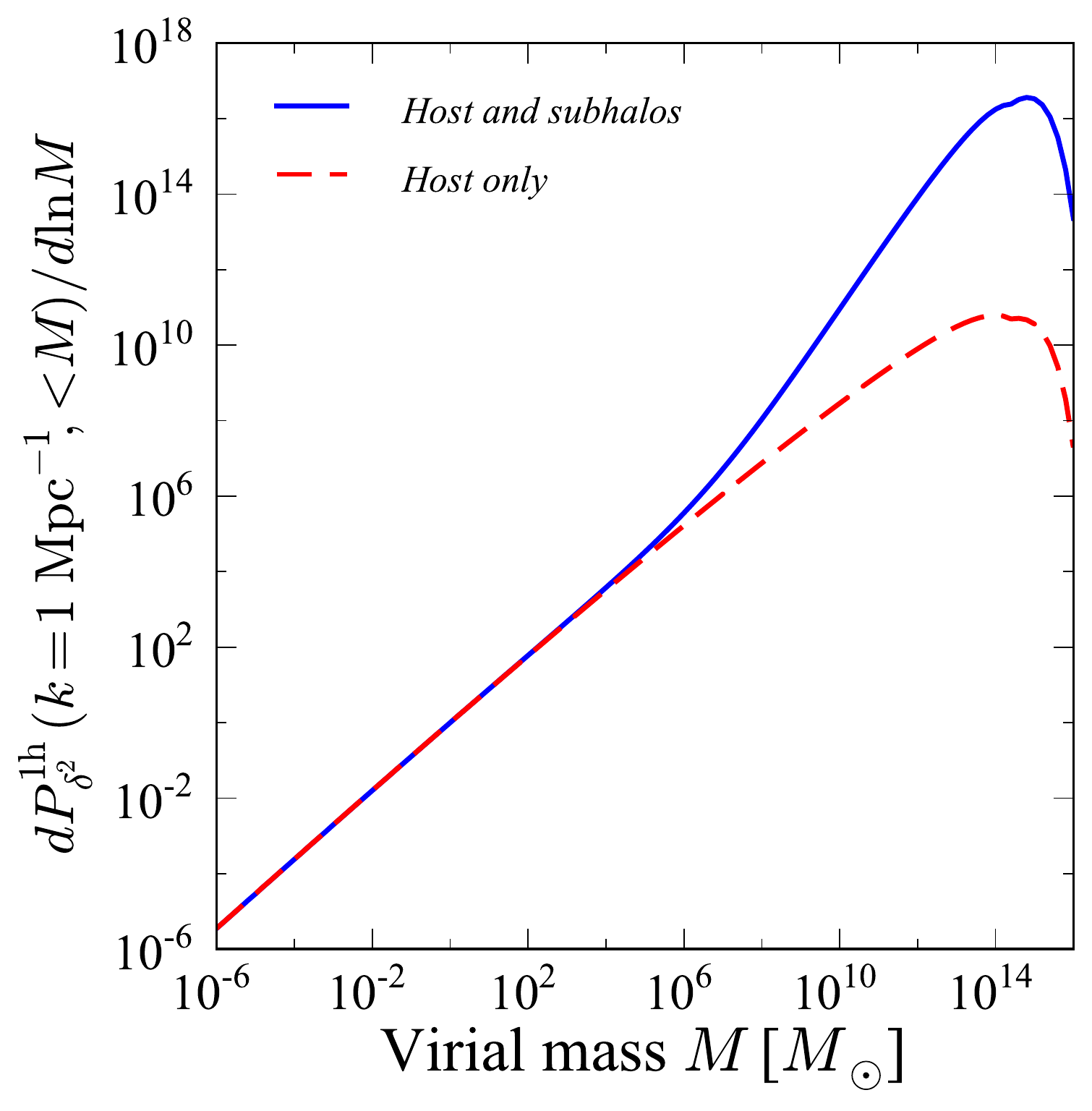}
  \caption{The same as Fig.~\ref{fig:dPdM} for $k =
  1~\mathrm{Mpc}^{-1}$, but in comparison with the host-halo
  contribution.}
  \label{fig:dPdM_host}
 \end{center}
\end{figure}

Fourier transform of the density-squared profile,
$\tilde u(k|M)$, is  the sum of the density-squared
profiles of the host halo and subhalos weighed by a fractional
luminosity of each component:
\begin{equation}
 \tilde u(k|M) = \frac{\tilde u_{\rm host}(k|M) + b_{\rm sh}(M) \tilde
  u_{\rm sh}(k|M)}{1+b_{\rm sh}(M)}.
\label{eq:weightedsum}
\end{equation}
Here, we ignore a contribution from the cross term,  $2\rho_{\rm
host}(r)\rho_{\rm sh}(r)$, which is important only when $\rho_{\rm host}(r)
\sim \rho_{\rm sh}(r)$ at the same radius, $r$. Given that
spatial distributions of the host halo and subhalo contributions are
quite different (the host halo being important inside the scale radius
and the subhalos being important outside), this approximation is very
good. (See Appendix for the contribution of the cross term.)

Fourier transform of the host halo profile, $u_{\rm host}(r) \propto
\rho_{\rm host}^2(r)$, has an analytic form~\cite{Ando:2006}, but here
we use an even simpler fitting formula,
\begin{equation}
 \tilde u_{\rm host}(k|M) = \frac{1}{[1 + a (k r_s)^{2/b}]^b},
  \label{eq:Fourier fitting}
\end{equation}
with $a = 0.13$ and $b = 0.7$, which is largely independent of $c_{\rm
vir}$~\cite{Ando:2009a}.
On the other hand, we obtain the density-squared profile of subhalos,
$u_{\rm sh}(r|M)$, by deprojecting the surface brightness profiles of
numerical simulations~\cite{Gao:2012, Han:2012}, assuming spherical
symmetry as follows:
\begin{equation}
 u_{\rm sh}(r) \propto
  \left\{
   \begin{array}{ccc}
     \left[\left(\frac{r}{r_{200}}\right)^2 +
      \frac{1}{16}\right]^{-3/2}, & \mbox{for} & r \le r_{200}, \\
    \left(\frac{16}{17}\right)^{3/2}
     \left(\frac{r}{r_{200}}\right)^{-1}
     e^{-\eta (r/r_{200}-1)}, & \mbox{for} & r > r_{200},
   \end{array}
  \right.
\end{equation}
with $\eta = 2.78$.
Note that the distribution of subhalos is typically more
extended than the density profile of the host halo.
Its Fourier transform is
\begin{eqnarray}
 \tilde u_{\rm sh}(k|M) &=& A
  \left[\int_0^1 dx \frac{x^2}{(x^2 + 1/16)^{3/2}}
   \frac{\sin(\kappa x)}{\kappa x}
   \right. \nonumber \\ && {} \left.
  + \frac{64}{17^{3/2}}
  \frac{\kappa \cos\kappa  + \eta \sin\kappa}
  {\kappa (\kappa^2 + \eta^2)}\right],
\end{eqnarray}
where $\kappa \equiv k r_{200}$ and $A \approx 0.64$ is the normalization
constant such that $\tilde u_{\rm sh}(0) = 1$.

Figure~\ref{fig:uFourier} shows $\tilde u(k | M)$ for various host-halo
masses, $M$. When $\tilde u(k)$ is close to unity (for small $k$), the
halo can be regarded as a point source.  
On the other hand, when $\tilde u(k)$ deviates
significantly from unity at a given wave number $k$, 
the source extension cannot be ignored at that wave number, and
the power spectrum (especially the one-halo term) is suppressed.
Figure~\ref{fig:uFourier} shows that $\tilde u(k)$ is larger for
smaller host halos, which are less extended. It also shows that
the contributions from subhalos are more important for larger host
halos. 
As the distribution of subhalos is more extended than the
density profile of the host halo, the subhalo contribution dominates at
small $k$ and the host-halo dominates at large $k$. This makes
a hump at scales corresponding to the scale radius, $r_s$.
In other words, annihilation from the smooth host-halo
component dominates inside the scale radius, where subhalos are tidally
disrupted.

Figure~\ref{fig:dPdM} shows the integrand of the one-halo power
spectrum of $\delta^2$, $dP_{\delta^2}^{1h}(k, <M) / d\ln M$
[Eq.~(\ref{eq:1h})], at $z = 0$ for various wave numbers,
$k$. (Note that the lines are normalized at $M = 1~M_\odot$, and thus it
shows relative contributions rather than absolute.)
The bulk of the contributions come from large-mass halos, and halos
smaller than a typical dwarf size ($M < 10^6 M_\odot$) do not make any
sizable contributions to any relevant ranges of $k$.
This is particularly true for scales larger than a typical cluster size
(i.e., $k \alt 1 ~\mathrm{Mpc^{-1}}$).
For smaller scales, on the other hand, a relative importance of
large-mass halos is smaller, as large-mass halos are more extended and
thus the power from them is suppressed (as also shown in
Fig.~\ref{fig:uFourier}). 

Figure~\ref{fig:dPdM_host} is the same as Fig.~\ref{fig:dPdM} for $k =
1 ~ \mathrm{Mpc^{-1}}$, but the host-halo contribution is also shown. We
find that the impact of subhalos on the power spectrum
is much greater than that on the mean intensity (as shown in
Fig.~\ref{fig:dfdM_host}), and it can boost the power spectrum by
almost four orders of magnitude at this particular wave number.

\begin{figure}[t]
 \begin{center}
  \includegraphics[width=8.5cm]{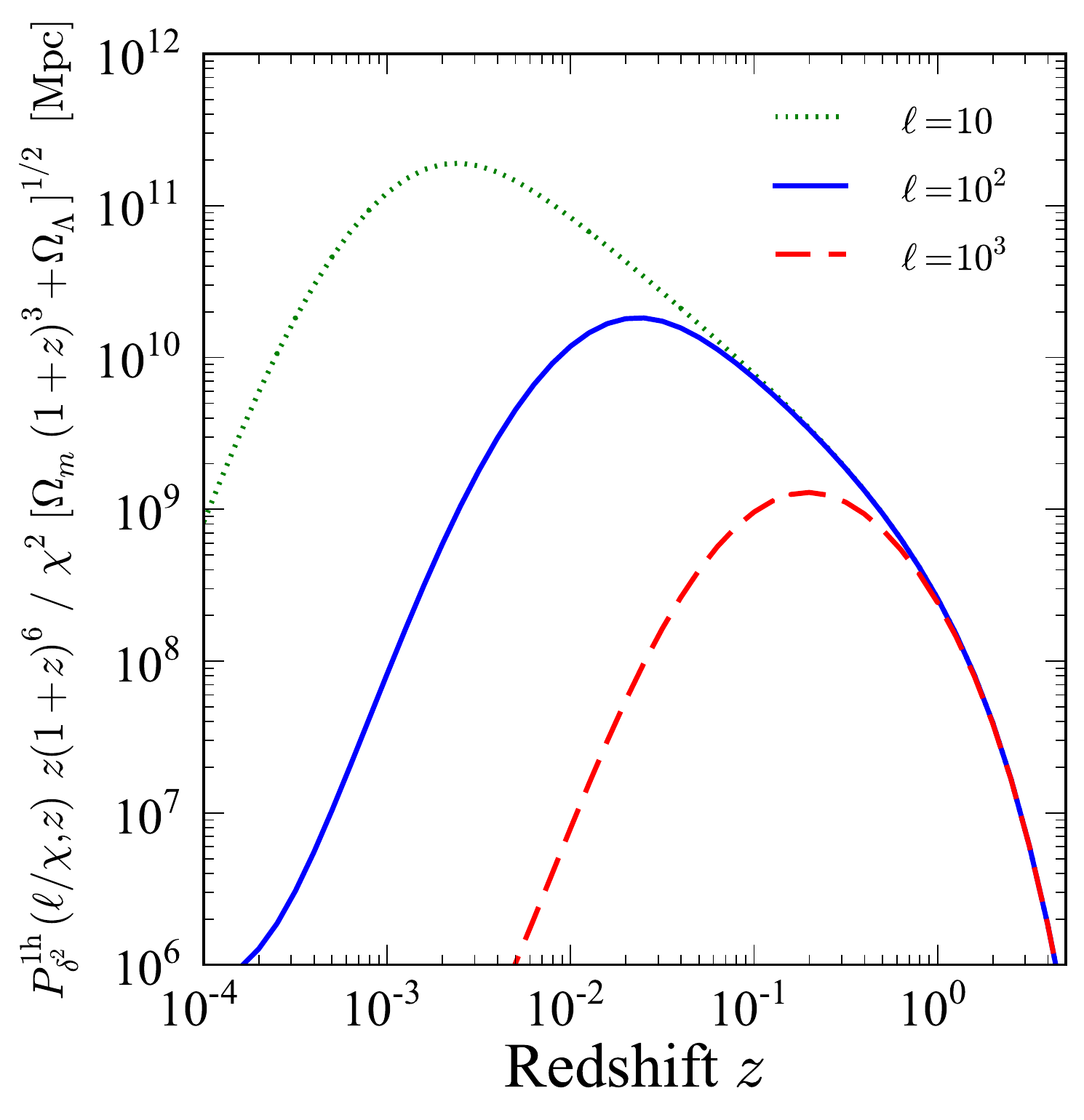}
  \caption{One-halo contributions to the angular power spectrum,
  $C^{\rm 1h}_\ell$, at $\ell=10$, 100, and 1000 as a function of redshifts.}
  \label{fig:ps1h_z}
 \end{center}
\end{figure}
\begin{figure}[t]
 \begin{center}
  \includegraphics[width=8.5cm]{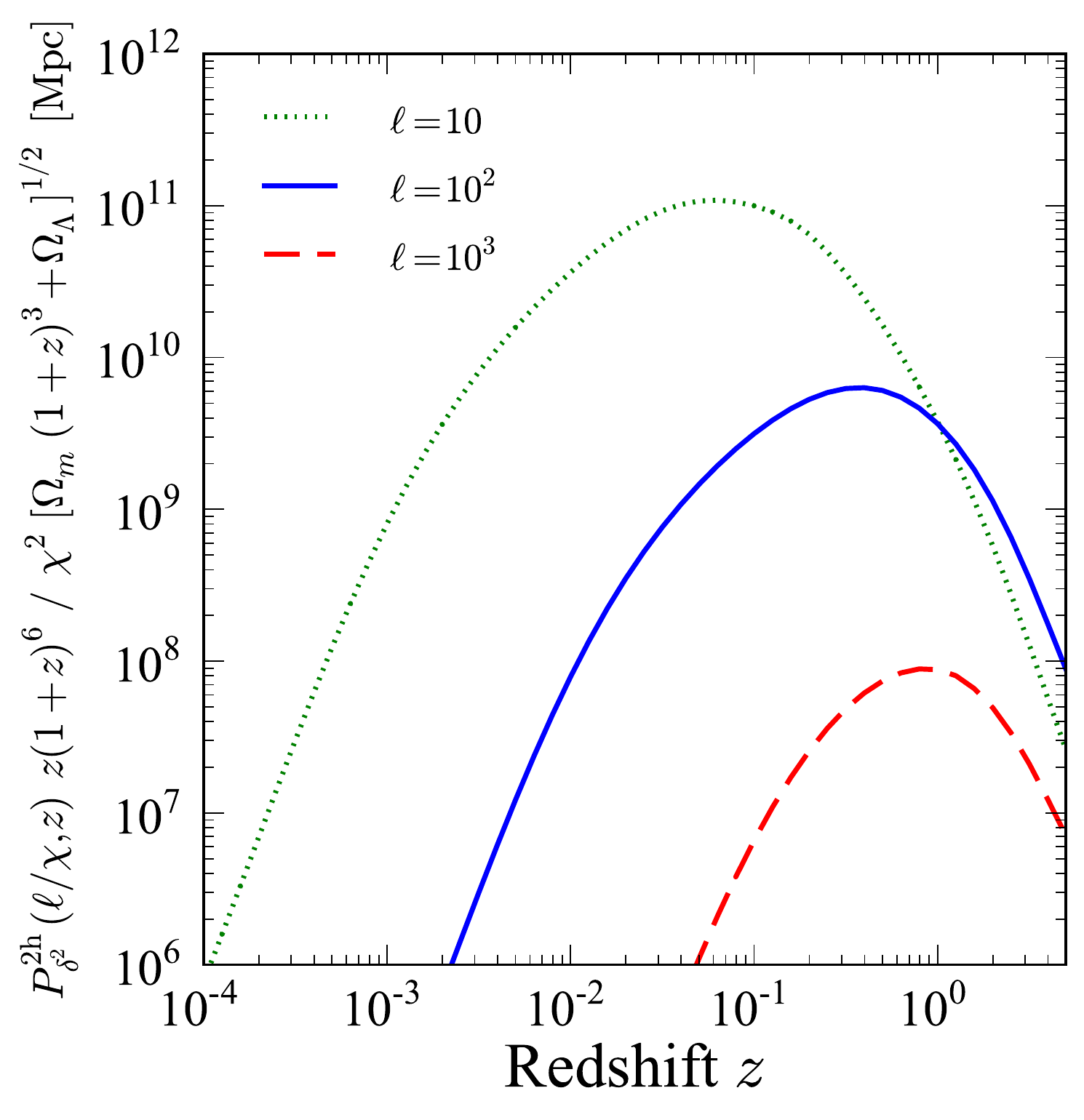}
  \caption{The same as Fig.~\ref{fig:ps1h_z} but for the two-halo
  contributions to the angular power spectrum, $C_\ell^{\rm 2h}$.}
  \label{fig:ps2h_z}
 \end{center}
\end{figure}

After projected on the sky, the three-dimensional wave number, $k$, and the
angular multipole, $\ell$, are related by $k = \ell /
\chi(z)$ for a given redshift, $z$ [see
Eq.~(\ref{eq:angular power spectrum})].
Note, however, that this simple relation is valid only for small
angular scales, $\ell\gg 1$~(e.g., \cite{Ando:2006, Ando:2007a}), on
which we mainly focus in this paper.

Figures~\ref{fig:ps1h_z} and \ref{fig:ps2h_z} show the
contributions to the angular power spectra at $\ell = 10$, 100, and 1000
from one-halo and two-halo terms, respectively, as a function of
redshifts.
To calculate a contribution to $C_\ell$ from a given $z$, we
multiply $P_{\delta^2}$ by some combination of functions of redshift
[see the integrand of Eq.~(\ref{eq:angular power spectrum}) and also the
redshift dependence in Eq.~(\ref{eq:window})].
We find that lower multipoles are dominated by nearby sources: one-halo
terms at $\ell=10$, 100, and 1000 are dominated by sources at $z\sim
0.002$, $0.02$, and $0.2$, respectively, whereas the dominant
contributions to the two-halo term come from somewhat higher redshifts.

One should not, however, include contributions from
arbitrarily small redshifts in the integral of Eq.~(\ref{eq:angular
power spectrum}), as cosmic variance in such small redshifts is so large
that taking the ensemble average (as we do here) no longer makes sense.
In addition, when a source is sufficiently close, it should give
enough gamma-ray fluxes to be identified as an individual source
which we can remove from the map.
In the following discussion, we use three different minimum
redshifts in the integration of Eq.~(\ref{eq:angular power spectrum}):
$z_{\rm min} = 0.001$, 0.003, and 0.01.
The r.m.s. overdensity within radii corresponding to these three redshifts
are 1.4, 0.74, and 0.28, respectively.
Since none of these are much larger than one, we can argue that our
results by setting these lower cutoffs are not subject to strong cosmic
variance. We also note that, as we show below, this choice does not make
any significant difference for the multipoles we consider ($\ell > 150$)
in this paper. 

\begin{figure}[t]
 \begin{center}
  \includegraphics[width=8.5cm]{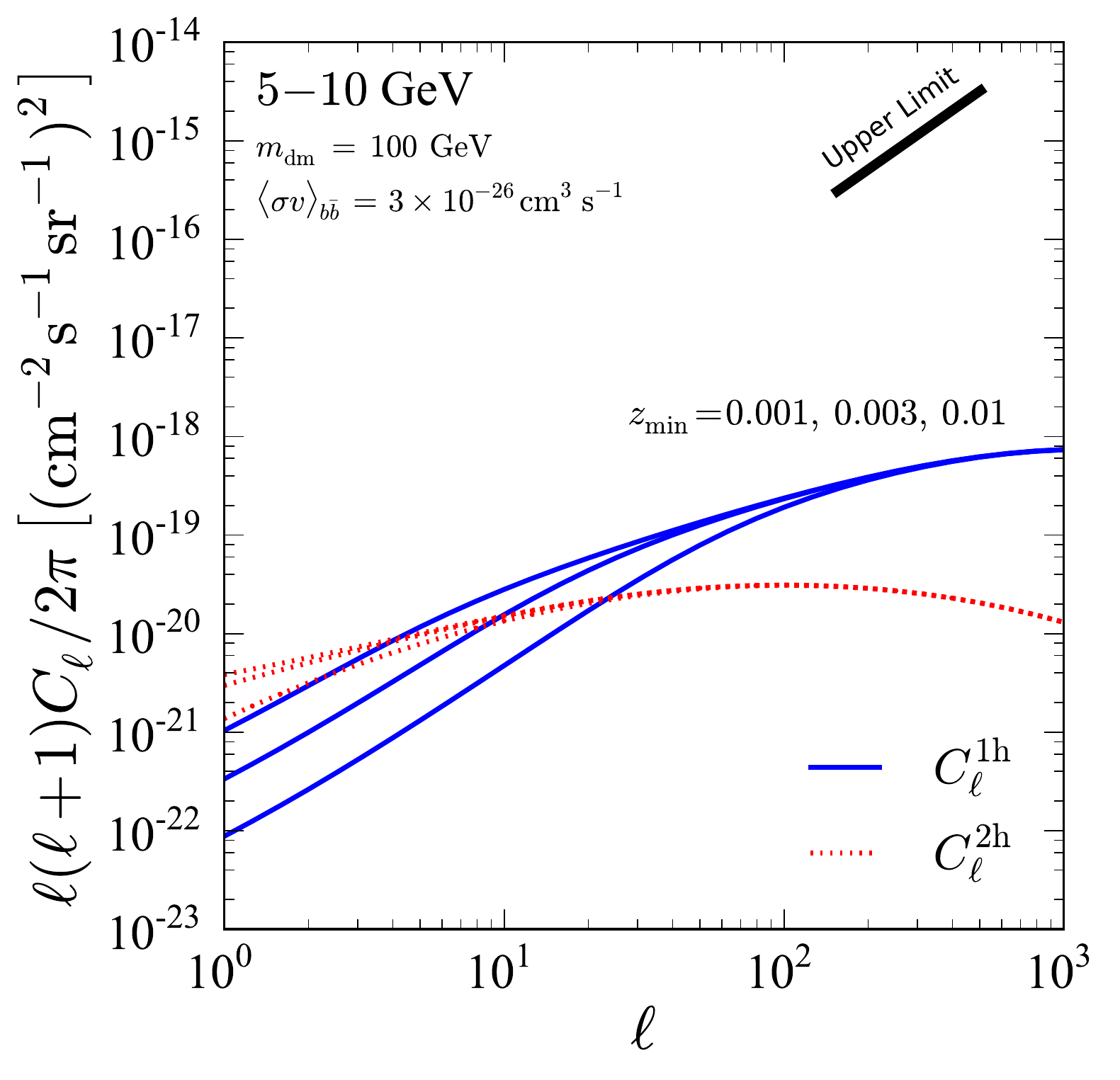}
  \caption{Predicted angular power spectrum
  of DGRB from dark matter annihilation in extragalactic halos,
  integrated over the energy range between
  5~GeV and 10~GeV. (This energy range is most sensitive to
  dark matter particles with $m_{\rm dm}=100$~GeV annihilating into
  $b\bar b$; see Fig.~\ref{fig:spectrum}.) The one-halo (solid) and
  two-halo (dotted) terms are
  shown separately, for three different minimum redshifts $z_{\rm
  min}=0.003$, 0.001, and 0.01 (from top to bottom lines). 
  The particle physics parameters are $\langle \sigma v \rangle = 3
  \times 10^{-26} ~ 
  \mathrm{cm^{3} ~ s^{-1}}$ and $m_{\rm dm} = 100 ~ {\rm GeV}$, and the
  $b\bar b$ annihilation channel is assumed. For comparison, the thick
  solid line shows the upper
  limit on the angular power spectrum in $155 \le \ell \le
  504$ from the 22-month data of Fermi-LAT~\cite{FermiAnisotropy} with
  the blazar contribution subtracted~\cite{Cuoco:2012}.}
  \label{fig:Cl}
 \end{center}
\end{figure}

In Fig.~\ref{fig:Cl}, we show both the one-halo and two-halo terms of
the angular power spectrum, $C_\ell$, integrated over 5--10~GeV energy
range, for three different values of $z_{\rm min}=0.001$, 0.003, and
0.01. 
The particle physics parameters are $\langle \sigma v \rangle = 3\times
10^{-26} ~\mathrm{cm^3 ~ s^{-1}}$ and $m_{\rm dm} = 100 ~ {\rm GeV}$,
and we assume the $b\bar b$ annihilation channel. Note that $C_\ell$
scales as $\langle \sigma v\rangle^2$. 
Taking smaller $z_{\rm min}$ increases the power at large angular
scales, in particular for the one-halo term, because of the
contributions from closer, more extended halos.
For the rest of this paper, we shall use $z_{\rm min} = 0.003$ for
definiteness.

\begin{figure}[t]
 \begin{center}
  \includegraphics[width=8.5cm]{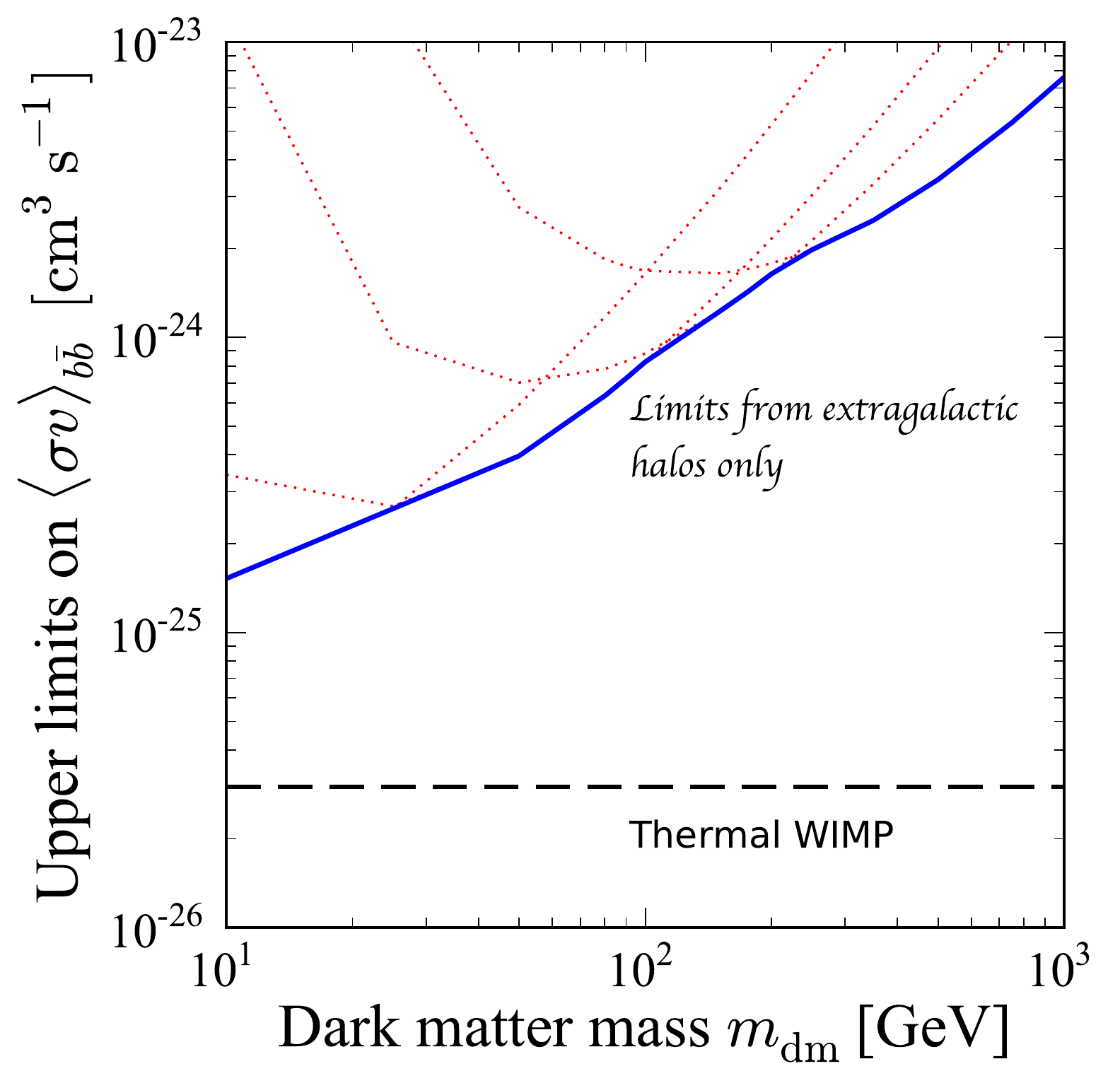}
  \caption{Upper limits on the annihilation cross section into the $b
  \bar b$ final state as a function of the dark matter masses, $m_{\rm
  dm}$. The limits are obtained from the measurement of the angular
  power spectrum at $\ell = 155$~\cite{FermiAnisotropy} with the
  blazar contribution subtracted~\cite{Cuoco:2012}, and the theoretical
  model of dark matter annihilation 
  from extragalactic halos. The dotted lines show the limits from different
  energy ranges: 1.04--1.99~GeV, 1.99--5~GeV, 5--10.4~GeV, and
  10.4--50~GeV from bottom to top at low-mass region. The solid line
  shows the combined limits, and the horizontal dashed line
  shows the canonical cross section for the thermal-freezeout scenario.}
  \label{fig:limit_sigmav_EGH}
 \end{center}
\end{figure}

The thick solid line in Fig.~\ref{fig:Cl} shows the upper limits on
the angular power spectrum~\cite{FermiAnisotropy} (with the blazar
contribution subtracted \cite{Cuoco:2012}) between $\ell = 155$ and 504 
for 5--10.4~GeV, $C_{155 \le \ell \le 504} \le 8 \times
10^{-20} ~ \mathrm{(cm^{-2} ~ s^{-1} ~ sr^{-1})^2 ~
sr}$.
The current upper limits in 5--10.4~GeV, whose energy region
is most sensitive to dark matter particles with $m_{\rm dm}=100$~GeV
annihilating into $b\bar b$ (see Fig.~\ref{fig:spectrum}), are three
orders of magnitude larger than the prediction with the canonical
particle physics parameters. Recalling $C_\ell\propto \langle \sigma
v\rangle^2$, we find an upper limit on the cross section of $\langle
\sigma v \rangle \lesssim 8 \times 10^{-25} ~ \mathrm{cm^3 ~ s^{-1}}$
for the 100-GeV dark matter annihilating into $b\bar b$.

The blazar-subtracted upper limits on the angular power spectrum
are available in several energy bands, 1.04--1.99~GeV, 1.99--5~GeV,
5--10.4~GeV, and 10.4--50~GeV~\cite{Cuoco:2012}.
In Fig.~\ref{fig:limit_sigmav_EGH}, we show the combined upper limits on the
annihilation cross section for the $b\bar b$ channel, $\langle \sigma v
\rangle_{b\bar b}$, as a function of the dark matter masses, $m_{\rm
dm}$, using all the available data on the power
spectrum. (Note that the model is still based only on the extragalactic
contribution.) We calculate the upper limit on the cross section such
that the predicted $C_\ell$ at $\ell=155$ (the lowest multipole at which
the measurement is reported) is equal to the $2\sigma$ upper limit
reported by Ref.~\cite{Cuoco:2012}. 
The limits from each energy range are shown separately as the dotted lines.
The combined limits shown as the solid line are simply the best of
the four limits at a given dark matter mass. A more optimal
analysis would improve these limits.

\section{Cross correlation between dark matter and blazars}
\label{sec:cross correlation}
\begin{figure}[t]
 \begin{center}
  \includegraphics[width=8.5cm]{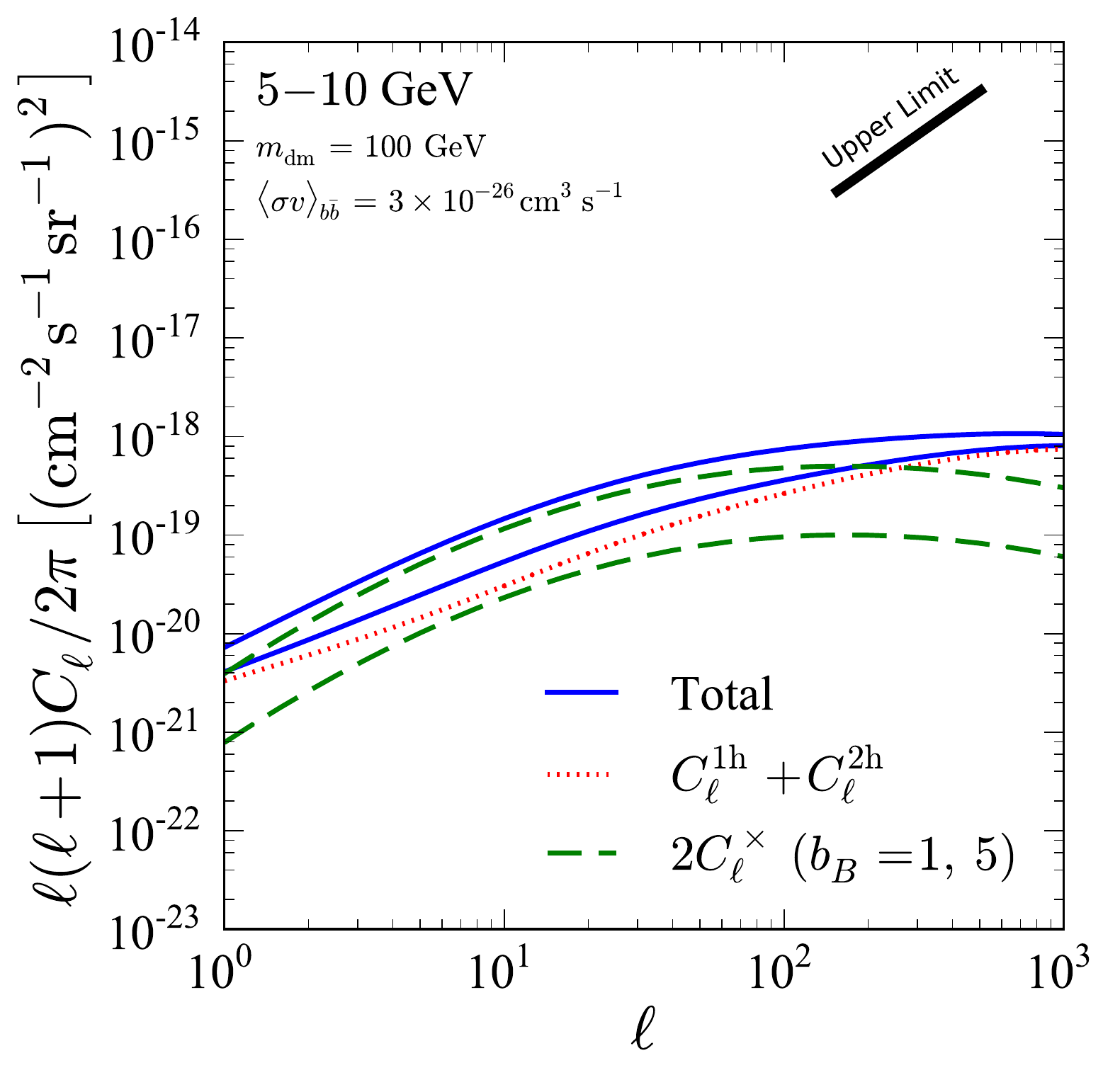}
  \caption{Predicted angular power spectra of DGRB in 5--10~GeV from dark
  matter annihilation only (dotted), dark matter--blazar
  cross correlation (dashed), and the sum of the two (solid). The
  particle-physics parameters are the same as those in  
  Fig.~\ref{fig:Cl}. For the blazar bias, we use $b_B = 1$ (bottom dashed/solid)
  and 5 (top dashed/solid).}
  \label{fig:Cl_cross}
 \end{center}
\end{figure}

As astrophysical gamma-ray sources, such as blazars associated with
supermassive black holes at the center of galaxies, reside in dark
matter halos, there is  a spatial correlation between gamma rays from
dark matter 
annihilation and blazars. The angular power spectrum of dark matter
annihilation including the cross correlation, which is equal to the
total power spectrum minus the blazar power spectrum, is given by
\begin{equation}
 C_\ell^{\rm dm} \equiv C_\ell-C_\ell^{\rm blazar}= C_\ell^{\rm 1h} +
  C_\ell^{\rm 2h} + 2 
  C_\ell^{\times},
\end{equation}
where $C_\ell^{\times}$ is the cross-power
spectrum computed from\footnote{We include  
blazars in the cross-correlation, ignoring other sources of
gamma rays. This may be justified, as blazars are so far known to be
the most dominant extragalactic gamma-ray sources in the GeV energy regime.
Other promising sources include star-forming
galaxies~\cite{Ando:2009b, Fields:2010, Makiya:2011, Lacki:2012}.} 
\begin{eqnarray}
 C_\ell^\times(E) &=& \int \frac{d\chi}{\chi^2} W_B([1+z]E, z)
  W_{\rm dm}([1+z]E, z)
  \nonumber\\ && {} \times
  P_\times \left(k = \frac{\ell}{\chi}, z\right),
  \label{eq:cross correlation}
\end{eqnarray}
where the subscripts ``$B$'' and ``dm'' denote blazars and dark matter,
respectively. 
Here, $W_{\rm dm}$ is the same as Eq.~(\ref{eq:window}), while $W_B$ is
the window function for the mean intensity of blazars [similarly
defined as Eq.~(\ref{eq:intensity}) but replacing $\langle \delta^2
\rangle$ with 1]:
\begin{equation}
 W_B([1+z]E, z) = \chi^2 \int_0^{\mathcal L(\mathcal F_{E, {\rm lim}}, z)}
  d\mathcal L~ \Phi_E(\mathcal L, z) \mathcal F_E(\mathcal L, z),
  \label{eq:window function for blazars}
\end{equation}
where $\mathcal L$ is the differential luminosity (i.e.,
luminosity per unit energy) at a given energy $E$,
$\mathcal F_E$ is the differential flux, and $\Phi_E (\mathcal L, z)$ is
the gamma-ray luminosity function of blazars.
The upper limit of the integral, $\mathcal L(\mathcal F_E, z)$, is the
luminosity giving the flux corresponding to the point-source sensitivity
of Fermi-LAT at a given redshift $z$, for which we adopt $10^{-8}
~\mathrm{cm^{-2} ~ s^{-1}}$ above 100~MeV.
More formal definitions of these quantities can be found in
Ref.~\cite{Ando:2007a}.

The three-dimensional cross power spectrum is
\begin{eqnarray}
 P_\times(k,z) &=&
  \langle b_B(z) \rangle
  \left[\left(\frac{1}{\Omega_m \rho_c}\right)^2
  \int dM \frac{dn}{dM}(M,z) \tilde u(k|M)
  \right. \nonumber\\&&{}\times \left.
  b_1(M,z)
  (1+b_{\rm sh}(M)) \int dV \rho_{\rm host}^2(r|M)\right]
  \nonumber\\&&{}\times
  P_{\rm lin}(k,z)
  \nonumber\\
 &=& \langle b_B(z) \rangle
  \sqrt{P_{\delta^2}^{\rm 2h}(k,z) P_{\rm lin}(k,z)},
  \label{eq:cross power}
\end{eqnarray}
where $\langle b_B(z) \rangle$ is the blazar bias averaged over the
luminosity function and the flux as follows:
\begin{equation}
 \langle b_B(z) \rangle =
  \frac{\int d\mathcal L \Phi_E(\mathcal L,z) \mathcal F_E(\mathcal L,z)
  b_B(\mathcal L,z)}{\int d\mathcal L \Phi_E(\mathcal L,z) \mathcal
  F_E(\mathcal L,z)},
\end{equation}
with the luminosity-dependent bias, $b_B(\mathcal L,z)$, and the same
upper and lower limits of integration as those in Eq.~(\ref{eq:window
function for blazars}).

Note that this power spectrum [Eq.~(\ref{eq:cross power})]
includes the two-halo term only. While the one-halo term, where dark
matter annihilation happens in the same halo that hosts a blazar, also
exists,  the previous study~\cite{Ando:2007a} shows that the one-halo
term of the cross correlation is much smaller than the two-halo term;
thus, we shall ignore the one-halo term of the cross-power spectrum.

For the luminosity function, $\Phi_E(\mathcal L, z)$, we adopt the
luminosity-dependent density-evolution (LDDE) model~\cite{Narumoto:2006,
Ando:2007b} with the gamma-ray spectra assumed to be a power law with an
index of $2.4$, which is in agreement with the spectrum of resolved
blazars as well as that of the DGRB.\footnote{More elaborated spectra in
combination with the luminosity function and the DGRB intensity are
studied in Refs~\cite{Inoue:2009,Abazajian:2011}.} 
Compared with the earlier study~\cite{Ando:2007b} where the luminosity
function was based on {\it pre}-Fermi data, we here adopt different
values for parameters of the luminosity function ($\kappa = 10^{-4}$, $q
= 3.5$, and $\gamma_1 = 1.05$) such that the model reproduces the
flux distribution of blazars resolved by
Fermi~\cite{FermiDiffuseSource}.

In Fig.~\ref{fig:spectrum}, we show the blazar contribution to the mean
intensity. It is difficult to explain the
 DGRB intensity measured by Fermi with blazars alone, in agreement with the
previous study~\cite{FermiDiffuseSource, Ajello:2012}.

Figure~\ref{fig:Cl_cross} shows the angular power spectra from the
cross correlation with $\langle b_B \rangle = 1$ and
5,  for the energy range between 5~GeV and 10~GeV.
The particle-physics parameters are the same as those in
Fig.~\ref{fig:Cl}.
We find that, if the annihilation cross section is around the
canonical value required to produce dark matter particles at the
right abundance by the thermal-freezeout mechanism, 
then the cross-correlation term cannot be ignored.
This is particularly important when the bias of blazars is
as high as 5.

\begin{figure}[t]
 \begin{center}
  \includegraphics[width=8.5cm]{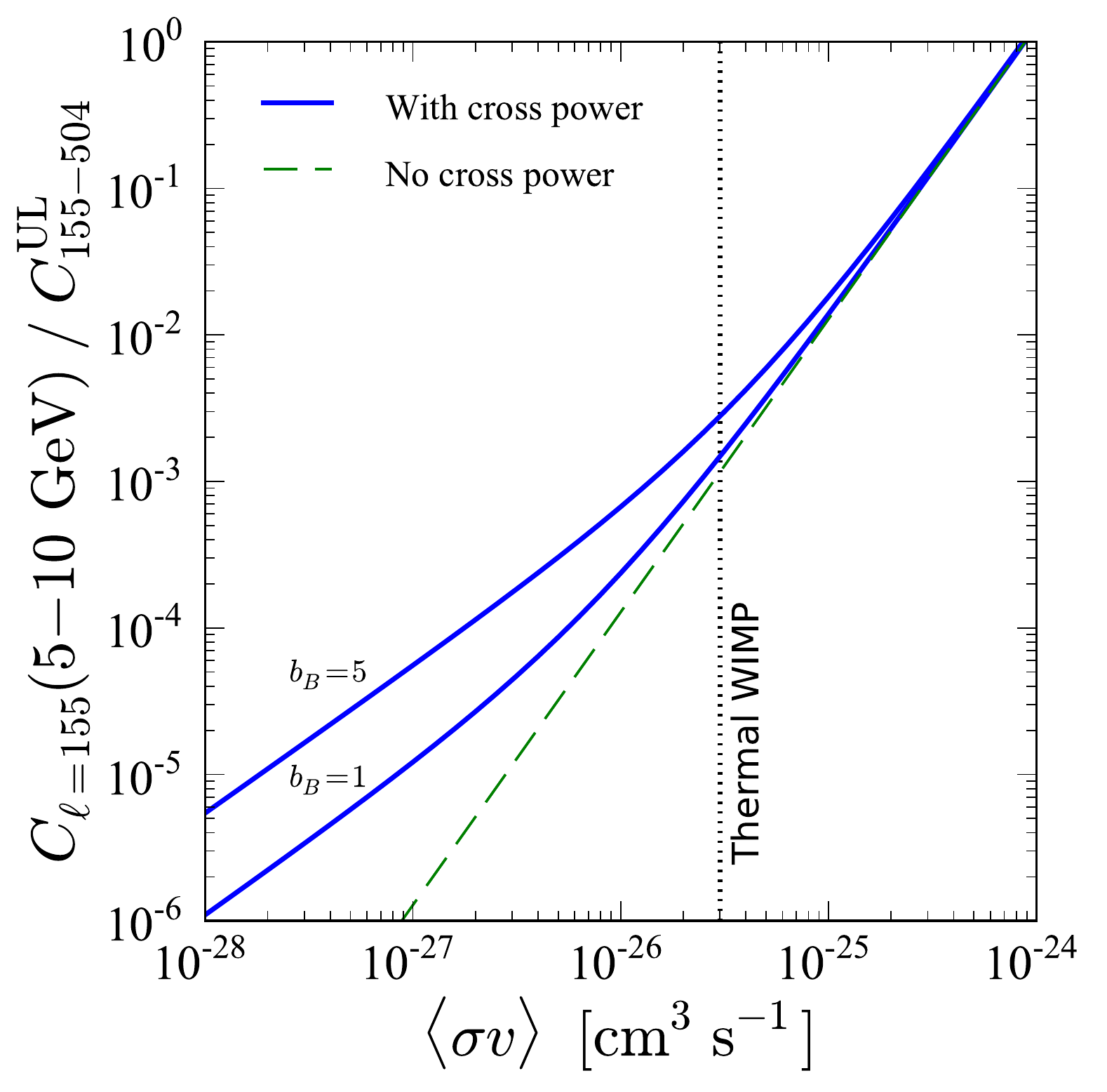}
  \caption{Predicted angular power spectrum at $\ell = 155$ in 5--10~GeV
  divided by the current upper limit, $C^{\rm UL}_{155 \le \ell
  \le 504}$, as a function of the annihilation cross section, $\langle
  \sigma v \rangle$. The solid lines include the dark-matter-blazar
  cross-correlation with $\langle b_B \rangle = 1$
  and 5, while the dashed line does not. 
  The vertical dotted line shows the canonical cross section for
  thermal WIMPs.}
  \label{fig:Cl_ratio_cross}
 \end{center}
\end{figure}

As the dark matter term ($C_\ell^{\rm 1h}+C_\ell^{\rm 2h}\propto \langle
\sigma v \rangle^2$) and the cross-correlation term
($C_\ell^\times\propto \langle \sigma v \rangle$) scale with $\langle \sigma v
\rangle$ differently, one may ask, ``At which value of the
annihilation cross section does the cross-correlation term become
important?''
Figure~\ref{fig:Cl_ratio_cross} shows $C_{\ell = 155}$ divided by
the upper limit, $C_{155 \le \ell \le 504}^{\rm UL}$, for 5--10~GeV. The
dashed line is without the cross correlation, while the solid lines are
with the cross correlation with bias of 1 and 5.
We find that the dark matter term dominates at  $\langle \sigma v
\rangle\gtrsim 3\times 10^{-26}~{\rm cm^3~s^{-1}}$, whereas the
cross-correlation term dominates at lower cross sections. 

\section{Galactic Contribution}
\label{sec:galactic}	
\begin{figure}[t]
 \begin{center}
  \includegraphics[width=8.5cm]{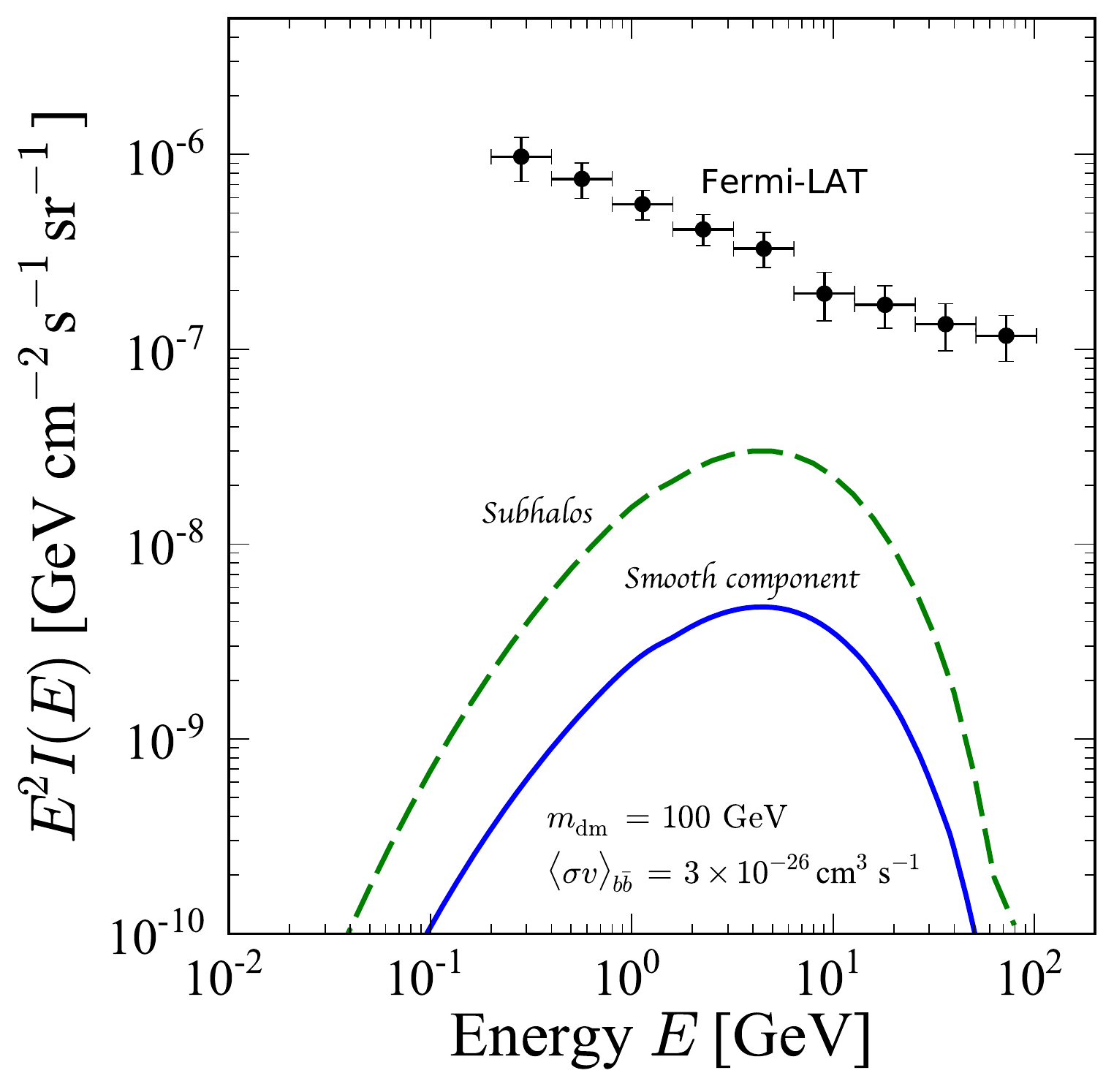}
  \caption{The same as Fig.~\ref{fig:spectrum}, but for the Galactic smooth
  component in the high Galactic latitude region outside of
  $|b|>30$~deg (solid). The dashed line is the same as that in
  Fig.~\ref{fig:spectrum}.}
  \label{fig:spectrum_smooth}
 \end{center}
\end{figure}
\begin{figure}[t]
 \begin{center}
  \includegraphics[width=8.5cm]{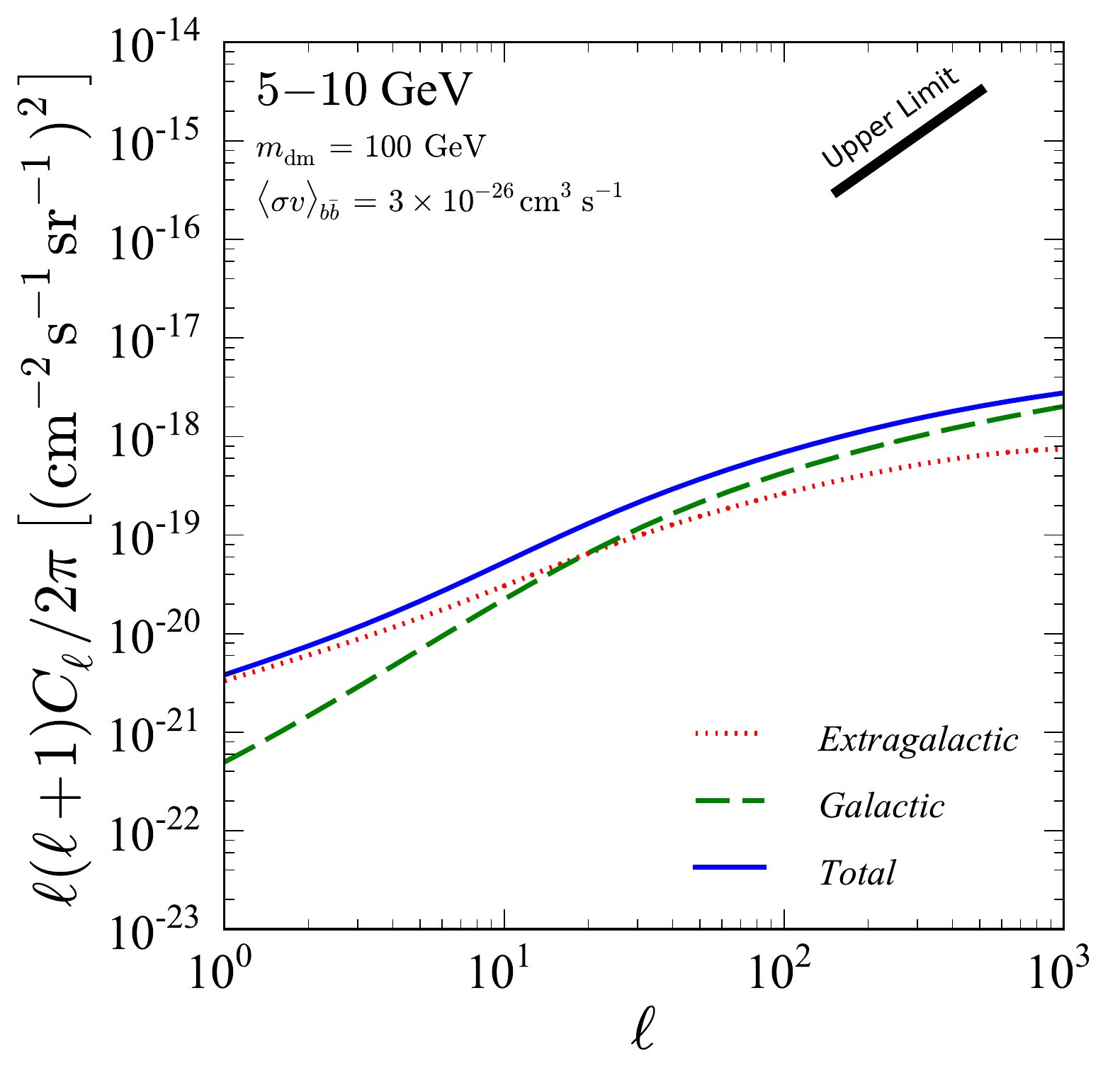}
  \caption{Predicted angular power spectra of DGRB in 5--10~GeV from dark matter
  annihilation in 
  extragalactic halos (dotted), Galactic subhalos (dashed), and the sum
  of the two (solid).}
  \label{fig:Cl_GSH}
 \end{center}
\end{figure}
\begin{figure}[t]
 \begin{center}
  \includegraphics[width=8.5cm]{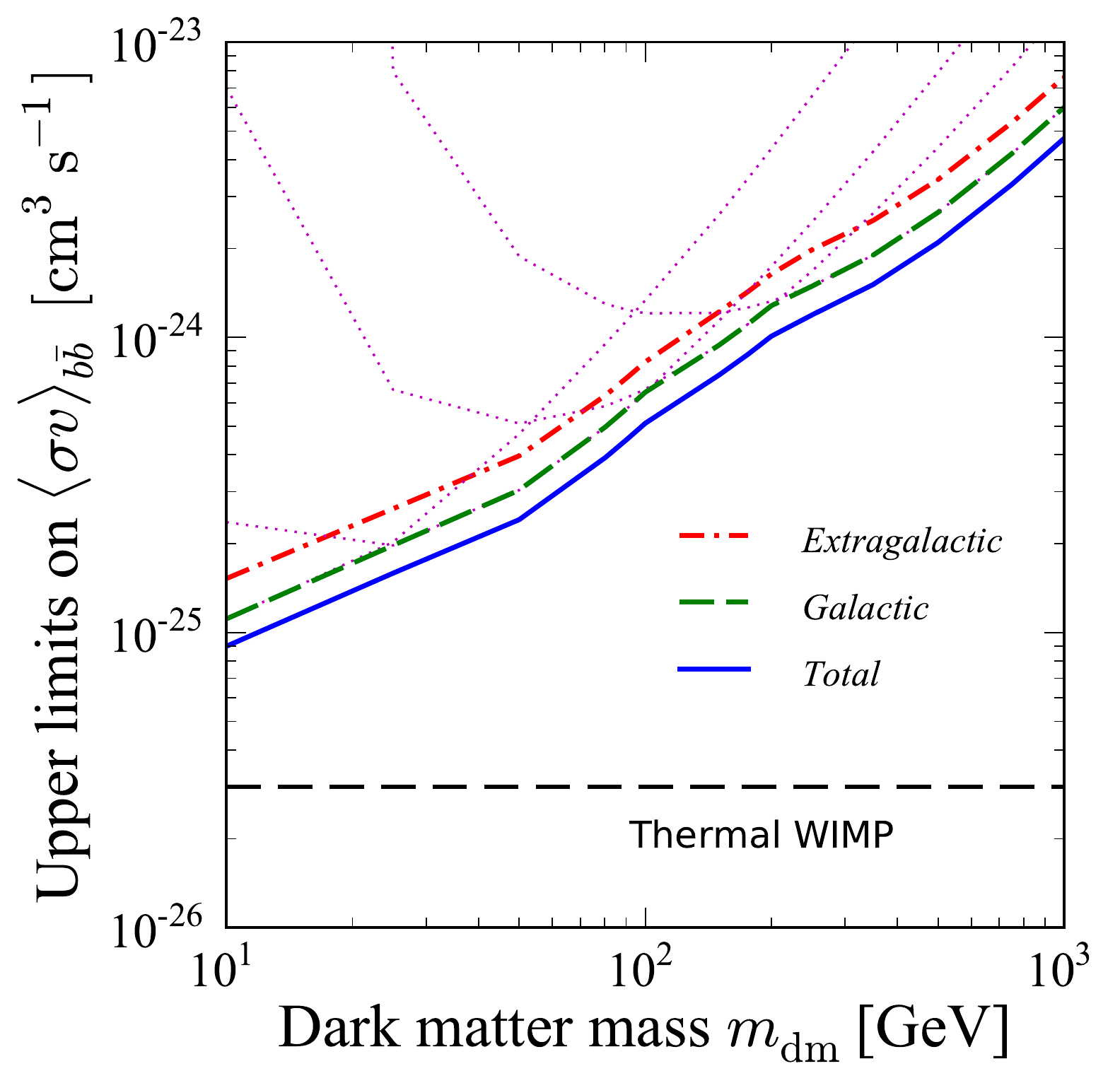}
  \caption{The same as Fig.~\ref{fig:limit_sigmav_EGH}, but
  for the limits obtained from the Galactic subhalos
  (dashed); extragalactic halos (dot-dashed); and the sum of the
  two (solid). The dot-dashed line is the same as the solid line in
  Fig.~\ref{fig:limit_sigmav_EGH}. The dotted lines show the
  Galactic-subhalo limits from each of four energy bins.}
  \label{fig:limit_sigmav_GSH}
 \end{center}
\end{figure}

Annihilation signals from subhalos in the Galactic halo containing the
Milky Way are typically comparable to, or even greater  than, the
extragalactic contribution~\cite{Oda:2005,
Pieri:2008,SiegalGaskins:2008, Fornasa:2009, Ando:2009a}; thus, we must
also take the Galactic subhalo contribution into account. We shall follow an
analytic treatment presented in Ref.~\cite{Ando:2009a}.

On the other hand, we do not include the contribution from a smooth
density profile of the host halo of the Milky Way in our
calculation. In the anisotropy analysis of the Fermi-LAT
data~\cite{FermiAnisotropy}, the low Galactic latitude region of $|b|\le
30$~deg is masked. Using a density profile of the smooth Galactic
component of Ref.~\cite{Ando:2009a}, we find that this mask brings the
smooth contribution to the mean intensity
down to about 10\% of the subhalo contribution, as shown in
Fig.~\ref{fig:spectrum_smooth}. As the smooth component does not have
much power on small angular scales, it can be safely ignored for
$\ell\gtrsim 100$ (but it can be comparable to the subhalo and
extragalactic contributions on large angular scales, $\ell\lesssim 10$).

The {\it angle-averaged} mean intensity from dark matter annihilation in
Galactic subhalos can be computed from
\begin{equation}
 I_{\rm sh} = \int dL \int_{s_\ast(L)}^{r_{\rm vir, MW}} ds
  \frac{d\overline{n_{\rm sh}}(L, s)}{dL} L,
\end{equation}
where $L$ is the gamma-ray luminosity of a subhalo, $s$ is the
line-of-sight coordinates, and $d\overline{n_{\rm sh}} / dL$ is the
angle-averaged luminosity function of subhalos.
The lower limit of the line-of-sight integral, $s_\ast (L)$, corresponds
to the flux sensitivity of Fermi-LAT, i.e., $L = 4\pi s_\ast^2 F_{\rm
sens}$.

The subhalo luminosity, $L$, is related to the subhalo mass, $M$, via
\begin{equation}
 L = B_{\rm sh}\frac{\langle \sigma v \rangle N_{\gamma, {\rm
  ann}}}{2m_{\rm dm}^2}
  \int dV_{\rm sh}~ \rho^2_{\rm sh}(r_{\rm sh}| M),
  \label{eq:subhalo luminosity}
\end{equation}
where $r_{\rm vir, MW}$ is the virial radius of the Galactic halo, the
subscripts ``sh'' denote subhalos, and $B_{\rm sh}$ is a
boost factor due to the presence of substructure in subhalos
(sub-subhalos).

We assume that the density profile of subhalos, $\rho_{\rm sh}$, is
well described by the NFW function [Eq.~(\ref{eq:NFW})].
Then, the volume integral of the density-squared has the analytic form
given by Eq.~(\ref{eq:volume integral of density squared}) with the
concentration parameter, $c_{\rm vir}$, replaced with $c_{\rm cut}$ that
corresponds to the cutoff radius of subhalos, i.e., $c_{\rm cut} \equiv r_{\rm
cut} / r_s$.
With this mass-luminosity relation [Eq.~(\ref{eq:subhalo
luminosity})] and the subhalo mass function $dn_{\rm sh} / dM$, one can
compute the luminosity function.
Most model inputs such as the subhalo mass function, spatial
distribution, and mass-concentration relation are adopted from recent
numerical simulations of the Galactic halo, {\it
Aquarius}~\cite{Springel:2008}.
More details on how to apply these models to gamma-ray computations are
described in Ref.~\cite{Ando:2009a}.

The intensity angular power spectrum is
\begin{equation}
 C_\ell^{\rm sh} = \frac{1}{16\pi^2} \int dL \int \frac{ds}{s^2}
  L^2 \frac{d\overline{n_{\rm sh}}(L, s)}{dL}
  \left|\tilde u_{\rm sh}\left(\frac{\ell}{s}, M\right)\right|^2,
  \label{eq:subhalo power}
\end{equation}
where $\tilde u_{\rm sh}(k, M)$ is the Fourier transform of the
density-squared profile of the subhalo distribution, 
which is given by Eq.~(\ref{eq:Fourier fitting}) if the 
density distribution of subhalos follows an NFW profile.
Note that Eq.~(\ref{eq:subhalo power}) only includes
``one-subhalo'' term, where one correlates two points in one identical
subhalo.
There is, however, the two-subhalo term that correlates two points in
two distinct subhalos, but this term is much smaller than the
one-subhalo term at small angular scales~\cite{Ando:2009a}.

Figure~\ref{fig:Cl_GSH} shows the predicted angular power spectra from
Galactic subhalos and extragalactic halos (but not including the cross
correlation). 
We have used the canonical model of the Galactic subhalos given in
Ref.~\cite{Springel:2008}, which has the mass resolution of
about $4 \times 10^4~M_\odot$. We have extrapolated their result down to
the Earth-mass scale (model A1 of Ref.~\cite{Ando:2009a}).
The intensity power spectrum is about the same for both the
extragalactic and Galactic components, with the latter slightly larger
in the angular scales constrained by Fermi-LAT.

In Fig.~\ref{fig:limit_sigmav_GSH}, we show the limits on $\langle
\sigma v \rangle$ from the Fermi-LAT data, taking into account both the
extragalactic and Galactic terms.
As expected, the limits from either alone are similar, and the combined 
limits improve by a factor of two.
In particular, for low-mass dark matter particles, the combined limits
 are only a factor of three larger than the canonical cross section.
The limits are weaker for larger masses. 

While our limits are not yet as stringent as those obtained from
analyses of dwarf galaxies~\cite{FermiDwarf, GeringerSameth:2011} or
galaxy clusters~\cite{Ando:2012,Han:2012}, where the canonical cross section is
already excluded for low-mass ($\sim$10~GeV) dark matter particles,
they are not so far away (i.e., only a factor of three to four
worse). Also, our limits are derived in a completely different way: they
are based on the diffuse emission rather than on individual objects, and
they are based on anisotropy rather than on the mean intensity.
It is certainly encouraging that the first limits using the DGRB
anisotropy are already not so far away from the best limits.

\section{Conclusions}
\label{sec:conclusions}
In this paper, we have used
 the angular power spectrum of DGRB recently detected in the
 22-month data of Fermi-LAT \cite{FermiAnisotropy} to place limits on
 the annihilation cross section of dark matter particles as a function
 of dark matter masses.
As dark matter annihilation occurs in all cosmological halos and
subhalos, our model includes all the contributing terms in the extragalactic
halos, the Galactic subhalos, and the cross correlation between dark
matter annihilation and blazars. The smooth Galactic component is
predicted to be sub-dominant in the high Galactic region ($|b|>30$~deg)
and is ignored.

We have revised our earlier model of the extragalactic contribution by
including the results from recent numerical simulations of the subhalo
distribution~\cite{Gao:2012}. Combined with the model of the Galactic
subhalos of Ref.~\cite{Ando:2009a}, we find that the Galactic and
extragalactic contributions are
comparable to each other. The cross correlation with blazars is
important for annihilation cross sections smaller than the canonical
value ($\langle \sigma v \rangle\lesssim 3\times 10^{-26}~{\rm
cm^3~s^{-1}}$).

By comparing our model with the upper limit on the
non-blazar contribution to the angular power spectrum of
DGRB~\cite{Cuoco:2012}, we find upper limits on the annihilation
cross section as a function of dark matter masses as shown in
Fig.~\ref{fig:limit_sigmav_GSH}.
The current limit from anisotropy excludes regions of $\langle \sigma
v \rangle\gtrsim 10^{-25} ~ \mathrm{cm^3 ~ s^{-1}}$ at the dark matter
mass of 10~GeV, which is only a factor of three larger than the
canonical value. The limits are weaker for larger dark
matter masses.  The first limits from DGRB anisotropy that we find in this
paper are already competitive with the best limits in the literature.

Our limits will improve as Fermi collects more data. At the
same time, an improvement in the analysis can significantly improve our
limits. Currently, the angular power spectrum on large angular scales,
$\ell< 155$, is not used because of a potential contamination by
the Galactic foreground emission (such as pion decay). As the angular
power spectrum of DGRB from dark matter annihilation, $C_\ell$ (without
multiplying by $\ell^2$), rises towards low multipoles, including the
low-multipole data will significantly improve the limits. This line of
investigation (i.e., a better characterization and removal of the
Galactic foreground) should be pursued.

\acknowledgments

The work of S.A. was supported by GRAPPA Institute at University of
Amsterdam and by NWO through Vidi Grant.

\appendix
\section{Effect of host-subhalo cross-term}

\begin{figure}[t]
 \begin{center}
  \includegraphics[width=8.5cm]{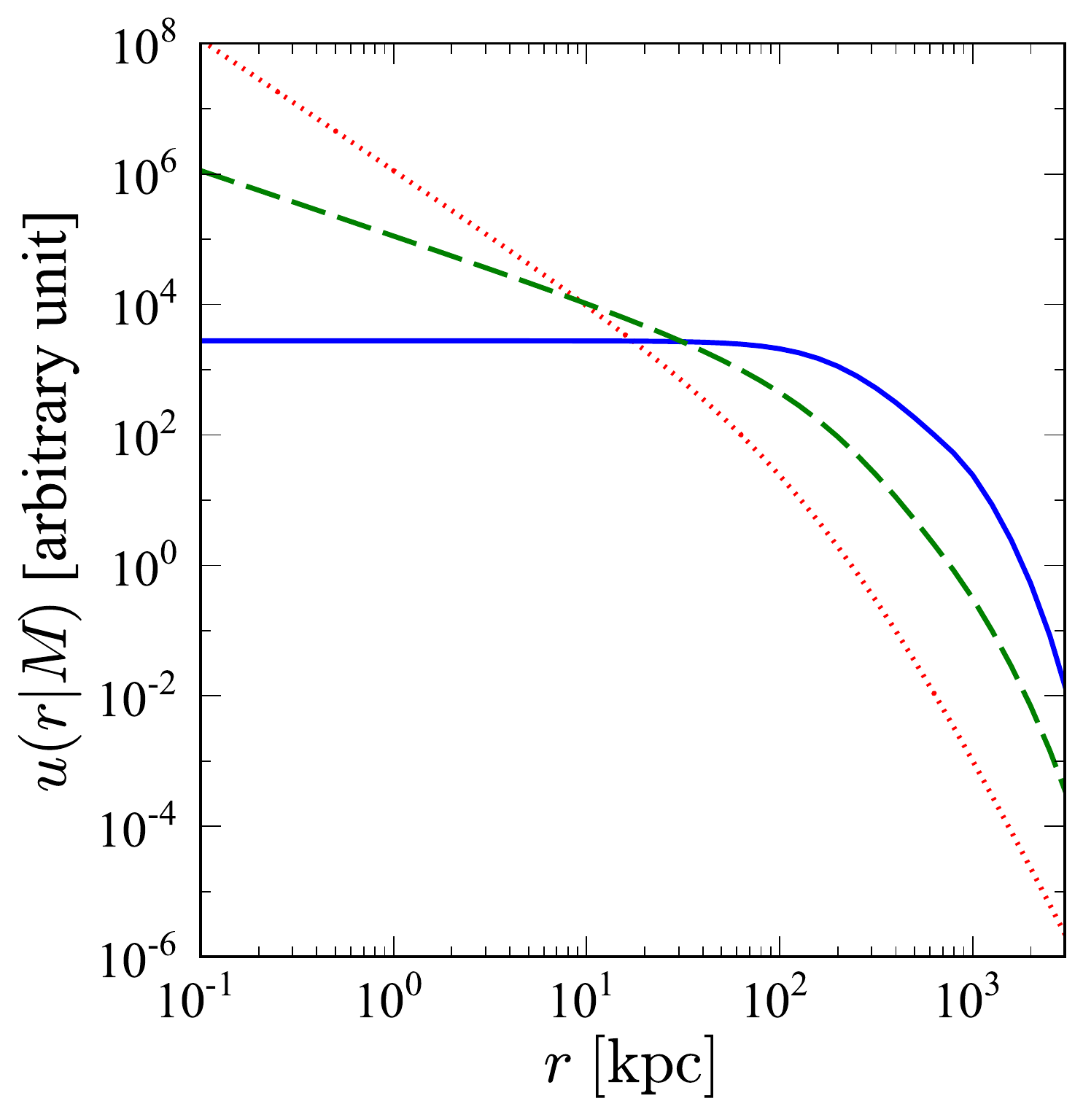}
  \caption{Density-squared profiles of a host halo (dotted); subhalos
  (solid); and a host-subhalo cross-term (dashed). The mass of the host
  halo is $M=10^{14}~M_\odot$ and the redshift is $z=0$. The virial
  radius is 1.2~Mpc and the scale radius is 210~kpc.}
  \label{fig:cross}
 \end{center}
\end{figure}
\begin{figure}[t]
 \begin{center}
  \includegraphics[width=8.5cm]{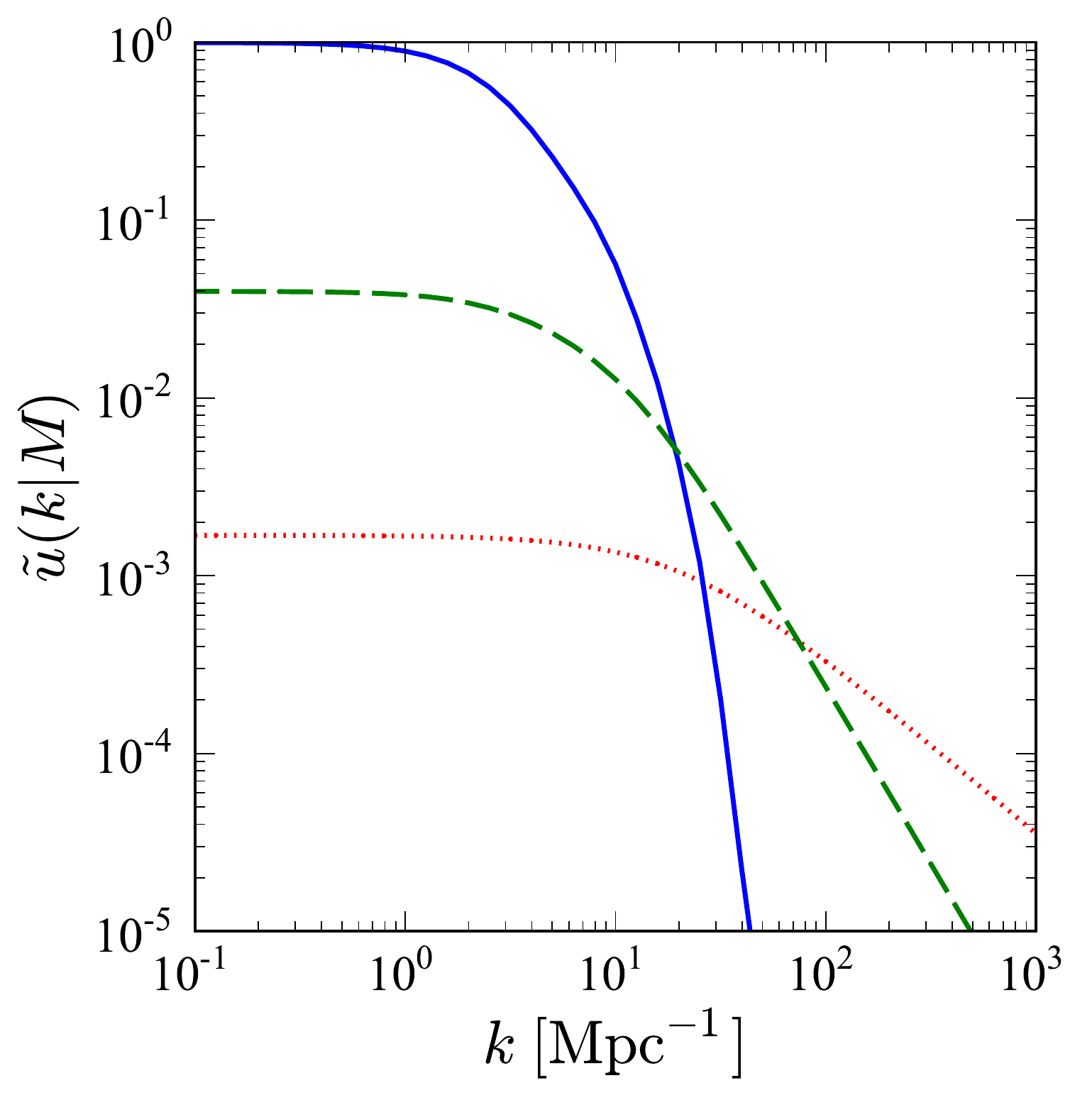}
  \caption{Fourier transform of the lines in Fig.~\ref{fig:cross}.}
  \label{fig:cross_k}
 \end{center}
\end{figure}

The dark matter annihilation signal is proportional to the density
squared. When both host halo and subhalo contributions are present, one
has the host-density-squared term, $\rho_{\rm host}^2$, the
subhalo-density-squared term, $\rho_{\rm sh}^2$, and the host-subhalo
cross term, $2\rho_{\rm host}\rho_{\rm sh}$. In our analysis, we have
ignored the cross term [see Eq.~(\ref{eq:weightedsum})], as
spatial distributions of the host halo and subhalo contributions are
quite different (the host halo being important inside the scale radius
and the subhalos being important outside). In this Appendix, we quantify
the importance of the cross term.

Figure~\ref{fig:cross} shows the density-squared profiles of a host
halo, subhalos, and the cross term. As expected, the cross term becomes
comparable to the other terms only within a narrow window in radii. For
the host halo mass of $M=10^{14}~M_\odot$ and the redshift of $z=0$, the
cross term becomes comparable to the other terms at $r\sim 20$~kpc,
which is $1/10$ of the scale radius,
$r_s=210$~kpc. Fig.~\ref{fig:cross_k} shows the Fourier transform.

\end{document}